\documentclass[sigconf]{acmart}

\usepackage{booktabs} % For formal tables
\usepackage{multirow} 
\usepackage[ruled, linesnumbered, vlined]{algorithm2e}
\usepackage[noend]{algorithmic}
\usepackage{balance}

\SetKwInOut{Input}{Input}
\SetKwInOut{Output}{Output}
\SetKwInOut{Variables}{Variables}
\SetKwComment{comm}{$\triangleright$\ }{}

\renewcommand\footnotetextcopyrightpermission[1]{} % removes footnote with conference information in first column
\newcommand{\hide}[1]{}
\newcommand{\eat}[1]{}

\newcommand{\nikos}[1]{[\underline{nikos}: \textcolor{red}{#1}]}

\begin{document}
%\title{Plane-Sweep based Evaluation of Interval Semi-Joins}
\title{Parallel In-Memory Evaluation of Spatial Joins}
\titlenote{Extended version of the ACM SIGSPATIAL'19 paper under the same title.}
%\titlenote{Produces the permission block, and copyright information}
%\subtitle{Extended Abstract}
%\subtitlenote{The full version of the author's guide is available as
%  \texttt{acmart.pdf} document}

%\author{Dimitrios Tsitsigkos$^\dagger$, Panagiotis Bouros$^\ddagger$, and Nikos Mamoulis$^\S$}
%\authornote{Dr.~Trovato insisted his name be first.}
%\affiliation{%
  %\institution{Department of Computer Science}
  %\institution{Aarhus University, Denmark}
%  \institution{Information Management Systems Institute, Athena RIIT,
%    Athens, Greece}
%  \institution{Institute of Computer Science, JGU, Mainz, Germany}
%  \institution{Dept. of Computer Science and Engineering, University of Ioannina, Ioannina, Greece}
%}
%\email{pbour@cs.au.dk}
%\email{bouros@uni-mainz.de}

\author{Dimitrios Tsitsigkos}
\affiliation{%
  \institution{Information Management Systems Institute}
  \institution{Athena RC, Athens, Greece}
%  \institution{Dept. of Computer Science and Engineering, University of Ioannina, Ioannina, Greece}
}
\email{dtsitsigkos@imis.athena-innovation.gr}

\author{Panagiotis Bouros}
\affiliation{%
  \institution{Institute of Computer Science}
  \institution{Johannes Gutenberg University Mainz, Germany}
%  \institution{Dept. of Computer Science and Engineering, University of Ioannina, Ioannina, Greece}
}
\email{bouros@uni-mainz.de}

\author{Nikos Mamoulis}
%\authornote{The secretary disavows any knowledge of this author's actions.}
\affiliation{%
  \institution{Department of Computer Science and Engineering}
  \institution{University of Ioannina, Greece}
}
\email{nikos@cs.uoi.gr}

\author{Manolis Terrovitis}
\affiliation{%
  \institution{Information Management Systems Institute}
  \institution{Athena RC, Athens, Greece}
%  \institution{Dept. of Computer Science and Engineering, University of Ioannina, Ioannina, Greece}
}
\email{mter@imis.athena-innovation.gr}

\renewcommand{\shorttitle}{Parallel In-Memory Evaluation of Spatial Joins}
\renewcommand{\shortauthors}{Tsitsigkos et al.}

\begin{abstract}
The spatial join is a popular operation in
spatial database systems and its
evaluation is a well-studied problem.
As main memories become bigger and faster and
commodity hardware supports parallel processing,
there is a need to revamp classic join algorithms which
have been designed for I/O-bound processing.
In view of this, we study the in-memory and parallel
evaluation of spatial joins, by re-designing
a classic partitioning-based algorithm to consider
alternative approaches for space partitioning\eat{,
and duplicate avoidance}.
Our study shows that, 
compared to a straightforward implementation of the algorithm,
our tuning can improve performance significantly.
We also show how to select appropriate
partitioning parameters based on data statistics, in order
to tune the algorithm for the given join inputs.
Our parallel implementation scales gracefully with the number
of threads reducing the cost of the join to at most one second even
for join inputs with tens of millions of rectangles. 
\end{abstract}

%
% % The code below should be generated by the tool at
% % http://dl.acm.org/ccs.cfm
% % Please copy and paste the code instead of the example below. 


\maketitle

\section{Introduction}\label{sec:intro}
The spatial join is a well-studied fundamental operation that finds
application in spatial database systems \cite{Guting94} and
Geographic Information Systems (GIS) \cite{Longley:2010:GIS:2544014}.
GIS, for example, typically store multiple thematic layers
(e.g., road network, hydrography), which are spatially joined in order to
find object pairs (e.g., roads and rivers) that intersect.
Spatial joins are also used
to support data mining operations such as clustering \cite{EsterKSX96} and pattern
detection \cite{CaoMC07}.

Given two collections $R$ and $S$ of spatial objects, the objective of
the {\em spatial intersection join} is
to find pairs $(r,s)$, such that $r\in R$, $s\in S$ and $r$ and $s$
have at least one common point.
Given the potentially complex geometry of the objects,
intersection joins are typically processed in two steps.
The {\em filter} step applies on spatial approximations of the
objects, typically {\em minimum bounding rectangles} (MBRs).
For each pair of object MBRs that intersect, the object geometries
are fetched and compared in a {\em refinement step}. 
%pbour
%In this paper,
Similar to the vast majority of previous work \cite{JacoxS07},
in this paper, we focus on the filter step.
%(in this paper) and consider refinement as the next step for future
%work.
Spatial intersection join algorithms can
also  be used for spatial distance joins between pointsets,
where the refinement step is trivial.
In particular, for an $\epsilon$-distance join,
each point $p$ is replaced (on-the-fly) by
a square of side $\epsilon$ centered at $p$;
for each pair of
intersecting squares, we can easily verify
if their distance is at most $\epsilon$.

A wide range of spatial join algorithms have been proposed in the
literature \cite{JacoxS07,BourosM19}.
Most of them assume that the input data are disk-based and their
objective is to minimize I/O accesses during the join.
Given the fact that
main memories become bigger and faster, 
main-memory join processing has recently received a lot of attendance
\cite{NobariQJ17}.
In addition, given that commodity hardware supports parallel
processing, multi-core join evaluation has also been the focus of
recent research.
Hence, in this paper, we focus on parallel in-memory evaluation of spatial joins on
modern hardware.

%We focus on the filter step of the spatial intersection join, hence we consider the join
%input to be two collections $R$ and $S$ of rectangles and the
%objective is to find the pairs of rectangles $(r,s), r\in R, s\in S$,
%such that $r$ and $s$ intersect.
Our focus is the optimization of
the simple, but powerful pa\-rti\-tio\-ning-based spatial join (PBSM) algorithm \cite{PatelD96}.
PBSM is shown to perform well in previous studies \cite{NobariQJ17}
and used by most distributed spatial data management systems
\cite{ZhangHLWX09,AjiWVLL0S13,EldawyM15}. In a nutshell, both
datasets are first partitioned using a regular grid; each tile (cell)
of the grid gets all rectangles that intersect it.
%The (possibly very
%large number of) tiles are grouped into a smaller number of partitions 
%in a round-robin fashion according to their z-ordering
%\cite{ZhangHLWX09}.
Each tile defines a smaller spatial join task. These tasks are
independent and can
be executed sequentially, assigned to different threads
or even to different machines in distributed evaluation.
The typical algorithm for processing each task is a plane sweep
algorithm based on forward scans \cite{BrinkhoffKS93}.
%is assigned to a different
%machine.
%We focus on the tile-to-tile join which is the core module of
%the entire process. That is, given a tile $c$ and the rectangle sets
%$R_c\subseteq R$ and $S_c\subseteq S$ that are assigned to $c$,
%compute the spatial intersection join between $R_c$ and $S_c$. 

\begin{figure}[htb]
\centering
  \includegraphics[width=0.45\columnwidth]{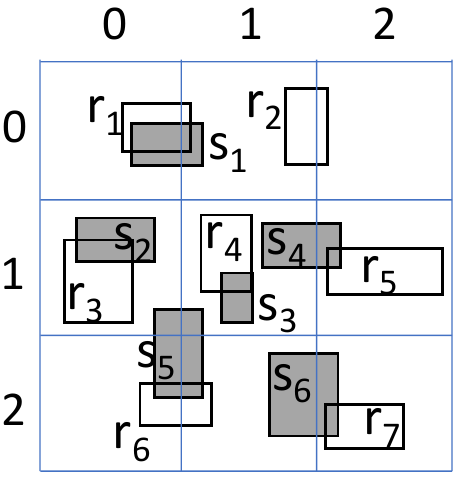}
  \caption{Example of PBSM}
  \label{fig:pbsm}
\end{figure}

As an example, consider the two sets of object MBRs
shown in Figure \ref{fig:pbsm}:
$R=\{r_1,r_2,\dots,r_7\}$ and $S=\{s_1,s_2,\dots,s_6\}$. 
By partitioning the rectangles
using a $3\times 3$ grid, we create 9 independent spatial join
tasks, one for each tile.
Note that some rectangles may be replicated
to multiple tiles, e.g., $s_1$ to tile (0,0) and (0,1).
Because of this, some pairs of rectangles may be found to intersect
each other in multiple tiles; e.g., pair $(r_1,s_1)$ intersect in
tiles (0,0) and (0,1). Outputting duplicate join pairs
can be avoided by reporting a pair of rectangles only if
a pre-determined reference point (typically, the top-left corner)
of the intersection region is in the tile \cite{DittrichS00}.
For example, $(r_1,s_1)$ is only reported by tile (0,0).

Most previous works do not focus on optimizing the performance of PBSM. 
In particular, there is no comprehensive study so far on how the number and type of partitions should be defined.
%We observed, by experimentation, 
%that changing the type and the number of partitions can make a big
%difference.
In this paper, we evaluate a 1D partitioning that divides the space into stripes, as opposed to the classic 2D partitioning, which uses a grid.
Further, we investigate, for each partition, the identification of
best direction of the sweep line.
%Moreover, we propose and
%evaluate an alternative approach for duplicate avoidance,
%which does not require the generation and corner checking
%of the redundant join pairs.
Finally, we show how both the
partitioning and the joining phases of the algorithm can be
parallelized.

Based on our experimental findings, the 1D partitioning results in 
a more efficient algorithm. 
In addition, increasing the number of partitions improves the performance of the algorithm, up to a point where adding more partitions starts having a negative effect.
%We also compare our duplicate avoidance technique with the method of 
%\cite{DittrichS00} and found them to have similar performance.
%\fix{find it to perform significantly faster in certain cases of partition-partition joining tasks.}
We present a number of empirical rules driven from data statistics (globally and locally for each partition) that can guide
  the selection of the algorithm's parameters.
Finally, we evaluate the performance of the parallel version of
the algorithm and show that it scales gracefully with the number of cores.
Overall, our optimized and scalable implementation is  
%can be  
orders of magnitude faster compared to a straightforward implementation.
For example, we can perform a
8.4M$\times$10M join in less than a second with 5 threads, which is 1-2 orders of magnitude improvement compared to the times for joining data of similar scale reported in recent work \cite{YuWS15,YouZG15,000100LCZG16,PavlovicHTKA16,SabekM17,PandeyKNK18}.

\eat{
Our study tries to answer the following questions:
\begin{itemize}
\item
Which is the best way to perform an in-memory join for two (small)
inputs? For this, we study how plane sweep algorithms perform.
\item
Can we improve the current practice of duplicate elimination \cite{DittrichS00}? Can we
avoid the generation of duplicate results overall?
\item
Can we further parallelize the tile-to-tile join computation? 
\end{itemize}
}

The rest of the paper is organized as follows. Section
\ref{sec:related} introduces the building
blocks of PBSM and presents related work.
Section \ref{sec:tuning} presents the directions along which we tune
the performance of the algorithm.
In Section \ref{sec:exp}, we present our evaluation on big real-world
datasets. Finally, Section \ref{sec:con} concludes the paper and gives
directions for future research.

\section{Background and Related Work}\label{sec:related}
%\pbour{this section is too long for essentially an experimental paper - the majority of the techniques are relevant are not tested, so the reader will lose his focus}
In this section, we review classic spatial join evaluation
approaches and more recent work for in-memory and distributed evaluation of
spatial joins.
%\subsection{General Join Evaluation Approach}\label{sec:smallmem}
In general, in order to join spatially two large object collections
$R$ and $S$, we first divide them into partitions which are small
enough  
and then join the partitions. 
We may also exploit an
existing partitioning or index. 
In either case, the join is broken down into numerous small problems
that can be solved fast in memory.
We first discuss how a (small) join problem can be processed
in memory, using a plane sweep algorithm.
Then, we review how data partitioning and indexing approaches can
be used to process bigger spatial join problems.
Finally, we review existing work in parallel and distributed spatial join evaluation.

\subsection{Evaluating Small Joins}\label{sec:ps}

For in-memory  processing of small spatial joins, 
a typical approach is to use adaptations of a plane sweep algorithm
that compute rectangle intersections \cite{PreparataS85}. 
The most commonly
used adaptation was suggested by Brinkhoff et al. \cite{BrinkhoffKS93}. 
Algorithm \ref{algo:fs} describes this method.
The join inputs $R$ and $S$ are first sorted 
based on their lowest value in one
dimension (e.g., $x_l$ of the $x$-dimension). 
Then, the sorted inputs are scanned concurrently
and merged as in a merge-join. 
This resembles a  line that (is perpendicular to and) sweeps along the sorting dimension.
For every value that the line encounters, say the lower $x$-endpoint
$r.x_l$ of a rectangle $r\in R$, the other input, i.e., $S$, is
{\em forwardly scanned} from the current rectangle $s'=s$,
while $s'.x_l$ is {\em not greater than}
the upper $x$-endpoint $r.x_u$ of $r$. 
All $s'\in S$ found in this scan
are guaranteed to $x$-intersect $r$, so for each of them
a $y$-intersection test is
applied (Line~8) to confirm whether $r$ and $s'$ intersect. 
Arge et al. \cite{ArgePRSV98} studied more classic (but
less simple to implement) versions of plane sweep based on
maintenance of {\em active lists} at every position of the sweep line, which have insignificant performance differences to Algorithm \ref{algo:fs}.

\begin{algorithm}[t]
 \LinesNumbered
  %\scriptsize
\Input{collections of rectangles $R$ and $S$}
\Output{set $J$ of all intersecting rectangles $(r,s) \in R\times S$}
\BlankLine
%$ J \leftarrow \emptyset$\;
\textbf{sort} $R$ and $S$ by lower $x$-endpoint $x_l$\;
$r \leftarrow$ first rectangle in $R$\;
$s \leftarrow$ first rectangle in $S$\;
\While{$R$ and $S$ not depleted}
{
%	\If(\comm*[f]{lock sweep line}){$r.\point{start} <
%	s.\point{start}$} 
	\If{$r.x_l < s.x_l$}
	{
		$s' \leftarrow s$\;
		%comm*{pivot in $S$}
		\While{$s' \ne$ null \textbf{\emph{and}} $r.x_u \geq s'.x_l$}
		{
			\If{$r.y$ intersects $s'.y$} %\comm*{test $y$-intersection}
			{
%				$J \leftarrow J~ \bigcup~ \{(r,s')\}$\comm*{add result}
				\textbf{output} $(r,s')$\comm*{update result}
			}
			$s' \leftarrow$ next rectangle in $S$\comm*{scan forward}
		}
		$r \leftarrow$ next rectangle in $R$\; %comm*{advance sweep line}
	}
	\Else
	{
		$r' \leftarrow r$\;
		%comm*{pivot in $R$}
		\While{$r' \ne$ null \textbf{\emph{and}} $s.x_u  \geq r'.x_l$}
		{
			\If{$r'.y$ intersects $s.y$} %\comm*{test $y$-intersection}
			{
%				$J \leftarrow J~ \bigcup~ \{(r',s)\}$\comm*{add result}
				\textbf{output} $(r',s)$\comm*{update result}
			}
			$r' \leftarrow$ next rectangle in $R$\comm*{scan forward}
		}
		$s \leftarrow$ next rectangle in $S$\; %comm*{advance sweep line}
	}
}
%\Return{$J$}
\caption{Forward Scan based Plane Sweep}
\label{algo:fs}
\end{algorithm}

\subsection{Data Partitioning}\label{sec:partitioning}

Spatially joining large inputs directly using Algorithm \ref{algo:fs}, without any preprocessing can be quite expensive. 
Some 20 years ago, the memories were too small to entirely fit the input data; hence, expensive sorting and sweeping would have to be performed in external memory.
Given this, {\em data partitioning} has been considered as a divide-and-conquer approach which splits the two inputs into smaller
subsets that can then be spatially joined fast in memory.
In a nutshell, each object collection is divided into
a number of partitions, such that objects that are spatially close to
each other fall in the same partition.
%based on the spatial proximity between the objects.
A partition from $R$ is then joined with a
partition from $S$ if their MBRs intersect.

A large number of spatial join algorithms that follow this paradigm have
been proposed.
They can be classified into {\em single-assignment, multi-join}
(SAMJ) methods and {\em multi-assignment,
single-join} (MASJ)  approaches \cite{LoR96}. 
SAMJ methods assign each object to exactly one partition; the
partitions are determined by spatial clustering heuristics.
A partition from one dataset (e.g., $R$) may
have to be joined with multiple partitions of the other dataset (e.g., $S$).
In MASJ, the borders of the partitions are pre-determined,
and an object is assigned to every partition it spatially intersects.
Each partition from $R$
is then joined with exactly one partition from
$S$ (which has exactly the same MBR).
Figure \ref{fig:partition}
shows the differences between these two partitioning schemes.
In SAMJ, illustrated in Figure \ref{fig:partition}(a),
the (dark grey) rectangles of dataset $R$
are divided to partitions $R_1$ and $R_2$, while the (hollow) rectangles
of $S$ are divided into groups $S_1$ and $S_2$. Partition $R_1$
only needs to be joined with $S_1$ because the MBR of $R_1$ does not
intersect the MBR of $S_2$.
However, $R_2$ should be joined with both $S_1$ and
$S_2$.
In MASJ, illustrated in Figure \ref{fig:partition}(b),
the datasets
are partitioned based on the space division defined by tiles $T_1$ and
$T_2$.
The rectangles from $R$ that intersect a tile (e.g., $T_1$) only have to 
be joined with the rectangles from $S$ that are assigned to the
same tile.
Note that objects $\{r_4,s_2,s_3\}$, which intersect both
tiles, are replicated.

\eat{
\begin{figure}
\centering
% For example, with the graphicx style use
\begin{tabular}{@{}cc@{}}
\includegraphics[width=0.4\columnwidth]{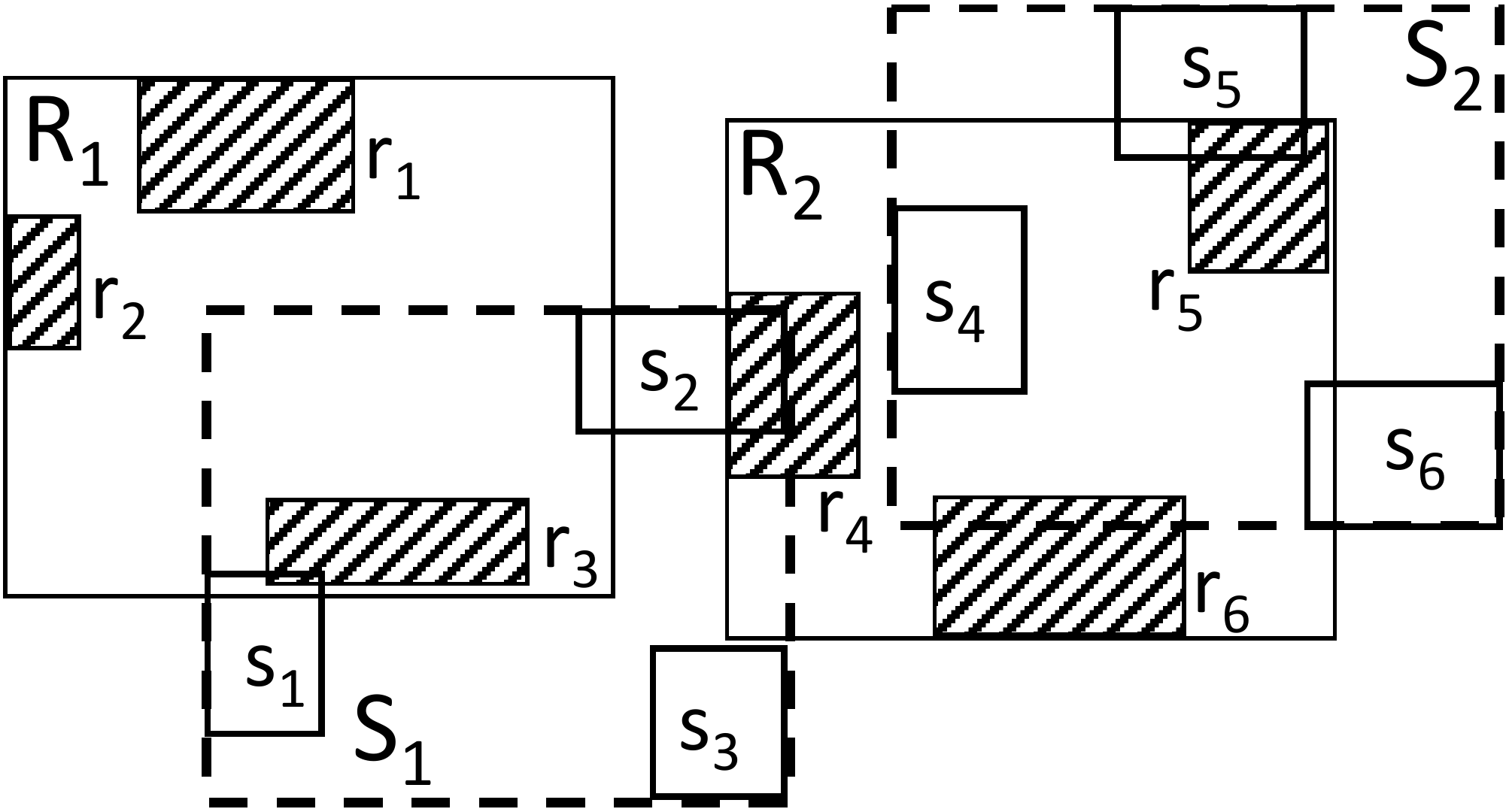}&
                                               \includegraphics[width=0.4\columnwidth]{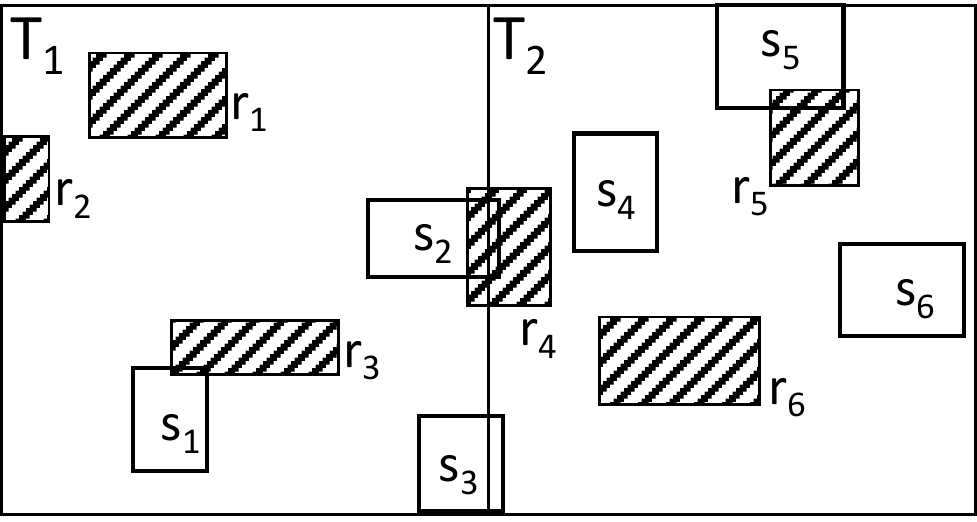} \\
(a) SAMJ&
                                          (b) MASJ\\
\end{tabular}
%
% If no graphics program available, insert a blank space i.e. use
%\picplace{5cm}{2cm} % Give the correct figure height and width in cm
%
\caption{Two classes of partitioning techniques}
\label{fig:partition}       % Give a unique label
\end{figure}
}
\begin{figure}
\centering
% For example, with the graphicx style use
\begin{tabular}{c}
\includegraphics[width=0.7\columnwidth]{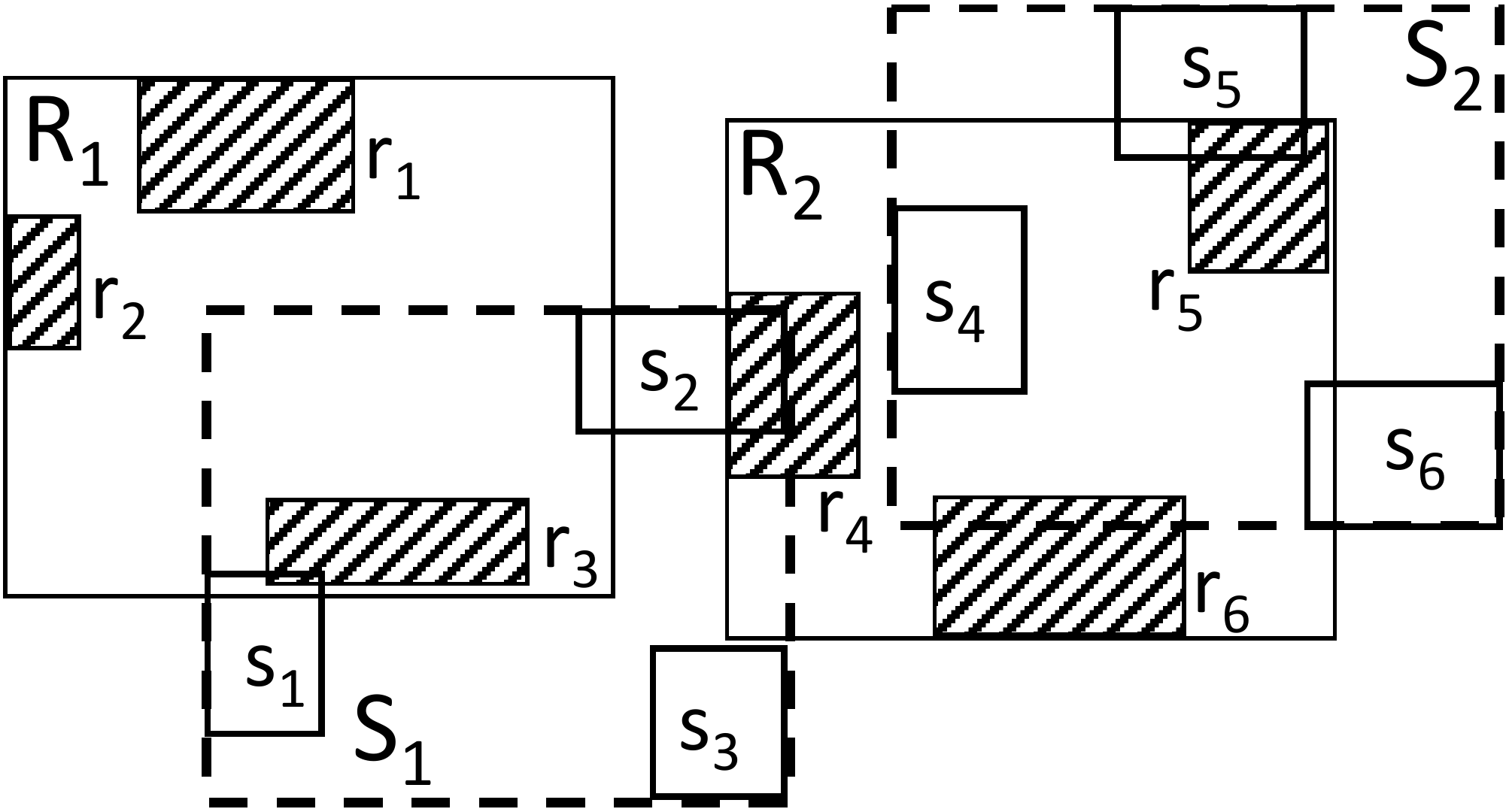}\\
(a) SAMJ\\\\
\includegraphics[width=0.7\columnwidth]{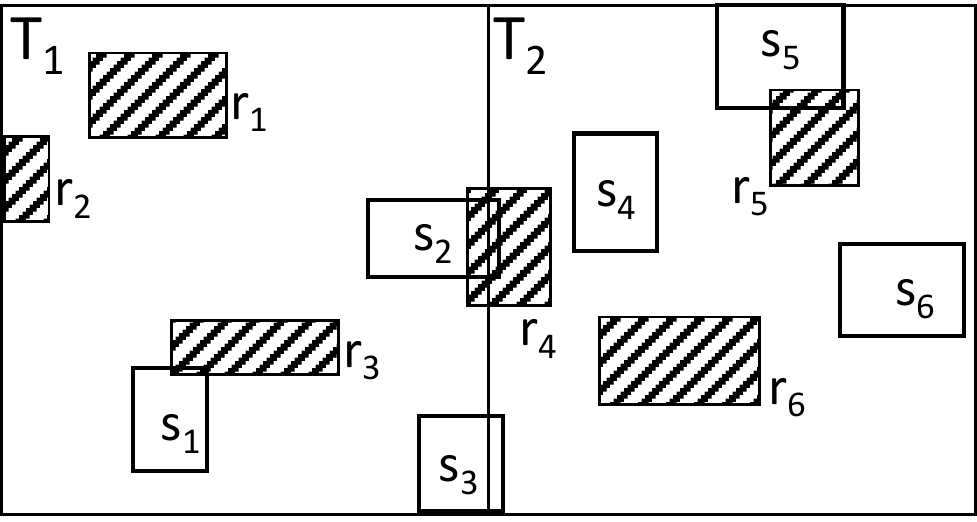} \\
(b) MASJ
\end{tabular}
%
% If no graphics program available, insert a blank space i.e. use
%\picplace{5cm}{2cm} % Give the correct figure height and width in cm
%
\caption{Two classes of partitioning techniques}
\label{fig:partition}       % Give a unique label
\end{figure}

A classic SAMJ approach, used when the two
inputs are indexed by R-trees \cite{Guttman84}, is the
{\em R-tree join} (RJ) algorithm of \cite{BrinkhoffKS93}. 
%takes as input two R-trees, which index $R$ and $S$, respectively.
RJ finds all pairs of entries $(e_R,e_S)$ one from each
root node of the trees that intersect. 
For each such pair, it
recursively applies the same procedure for the nodes pointed by $e_R$
and $e_S$, until pairs of leaf node entries (which correspond to
intersecting object MBRs) are found. 
For example, if $R_1$ and $R_2$ ($S_1$ and $S_2$, respectively) in Figure \ref{fig:partition}(a) 
are the two children of an R-tree root that indexes $R$ ($S$, respectively), then RJ would use
their MBRs to determine that  $R_1$ only needs to be joined with $S_1$.
%To join each pair of nodes, plane-sweep is used. RJ was later
%extended to a multiway join processing algorithm \cite{MamoulisP01}
%that applies to multiple R-trees.
Another SAMJ approach that does not rely on pre-defined indexes is
{\em Size Separation Spatial Join} \cite{KoudasS97}. 

The most popular MASJ approach is {\em Partition-based Spatial Merge Join}
(PBSM) \cite{PatelD96}. PBSM divides the space by a regular grid and
objects from both join inputs are assigned to all tiles which
spatially overlap them. 
For example, in Figure \ref{fig:partition}(b), $T_1$ and $T_2$ could
be tiles in a large rectangular grid that can be used to partition
both $R$ and $S$.
\eat{
  We may have a huge number of tiles which may have an uneven number of
objects.
In order to make the partitions as balanced as possible, we choose the
number of tiles to be much larger than the number of partitions and
then assign multiple tiles to each partition, such that the expected
set of objects per partition fits in memory and the partitions are as
balanced as possible. 
The assignment of tiles to partitions can be done in round-robin or 
by using a hash function. For example, assuming 3 partitions, tiles
$T_1, T_4, T_7$, etc. are assigned to partition 1,  $T_2, T_5, T_8$,
etc. to partition 2, etc.
}
%$In the join phase, f
For each partition, PBSM accesses the
objects from $R$, the objects from $S$ and performs their join
in memory (e.g., using plane-sweep). 
Since two replicated objects may intersect in
multiple tiles (e.g., see $r_4$ and
$s_2$ in Figure \ref{fig:partition}(b)), 
duplicate results may be produced. In order to avoid 
duplicates, a join result is reported by a tile only a pre-specified
reference point (e.g., the top-left corner) of the intersection region 
is in the tile \cite{DittrichS00}. 
%In Section \ref{sec:avoidance}, we
%propose an alternative mechanism, based on re-partitioning the contents of each tile, which avoids the production of duplicates.
Other MASJ approaches include {\em Spatial Hash Join} \cite{LoR96} and
{\em Scalable Sweeping-Based Spatial Join} \cite{ArgePRSV98}.

More recent spatial join algorithms consider the potential differences between the joined datasets in the distribution and density. Motivated by a neuroscience application, which requires joining datasets of contrasting density, Pavlovic et al. \cite{PavlovicTHA13} design a spatial join algorithm that partitions the dense dataset and `crawls' through the partitions guided by the object locations in the sparse dataset, skipping partitions that do produce any results.
Based on the same motivation,
a more sophisticated approach was proposed
in \cite{PavlovicHTKA16}, 
which adapts the type of partitioning (MASJ or SAMJ) and the join technique used locally, depending on differences in the densities of the two inputs.

\subsection{In-Memory Evaluation}\label{sec:smallmem}

Even with a large main memory that can accommodate the data, plane sweep can be too expensive if directly applied. The main reason behind this is that on a large map containing relatively small rectangles, the chances that two rectangles with intersecting $x$-projections
also intersect in the $y$-dimension are low.
Hence, plane sweep finds too many candidate pairs that $x$-intersect but do not materialize to actual results.

As a result, in-memory join approaches also consider data partitioning or indexing to accelerate processing.
For example, as in PBSM, 
a grid can be used to break the problem into
numerous small instances that can be solved fast. 
Algorithm {\em TOUCH} \cite{NobariTHKBA13} is an
effort in this direction, designed for scientific
applications that join huge datasets that have different density and
skew.
TOUCH first bulk-loads an R-tree for one of
the inputs using the STR technique \cite{LeuteneggerEL97}. 
Then, all objects from the second input are assigned to buckets corresponding
to the non-leaf nodes of the tree. 
Each object is hashed to the lowest tree node, whose MBR overlaps it,
but  no other nodes at the same tree level do.
Finally, each bucket is joined with the subtree rooted at the
corresponding node with the help of a dynamically created grid data
structure for the subtree. A recent
comparison of spatial join algorithms for in-memory data
\cite{NobariQJ17} shows that PBSM and TOUCH perform best 
and that the join cost depends on the data density and distribution.
Tauheed et al. \cite{TauheedHA15} suggest an analytical model for
configuring the grid of PBSM-like join processing in main memory;
however, this model (i) assumes a nested loops evaluation of each
partition-partition join and 
(ii) does not consider using the duplicate avoidance approach of \cite{DittrichS00}. 

\subsection{Parallel and Distributed Evaluation}
Early efforts on parallelizing spatial joins include extensions of the
R-tree join and PBSM algorithms in a distributed environment of
single-core processors with local storage. Consider the case of
joining two R-trees. Since overlapping pairs of root entries define
independent join processes for the corresponding sub-trees, these
tasks can be assigned to different processors
\cite{BrinkhoffKS96}. Two tasks may access a common sub-tree,
therefore a virtual global buffer is shared among processors to avoid
accessing the same data twice from the disk. An early approach in
parallelizing PBSM \cite{ZhouAT97} asks processors to perform the
partitioning of data to tiles independently and in parallel. Then,
each processor is assigned one partition and processors exchange data,
so that each one gets all objects that fall in its partition. The join
phase is finally performed in parallel.
%A {\em Partial Spatial Surrogate} approach \cite{PatelD00} replicates MBRs instead of entire tuples, while the exact object geometry is assigned to a single (home) node. In the join phase,  each operator looks at the fragments of the declustered datasets residing on its local disks and joins them using any centralized spatial join algorithm.

Recently, the research interest shifted to spatial join processing for
distributed cloud systems and multi-core processors. The
popularity and commercial success of the MapReduce framework motivated
the development of spatial join processing algorithms on clusters. 
The {\em Spatial Join with MapReduce} (SJMP) algorithm \cite{ZhangHLWX09}
is an adaptation of the PBSM algorithm in this direction. Initially, the space is
divided by a grid and tiles are grouped to partitions in a round-robin
fashion after considering them in a space-filling curve order.
During the {\em map} phase, each object is assigned to one or more
partitions based on the tiles it overlaps. Each partitioned object
also carries the set of tiles it intersects. Each partition
corresponds to a {\em reduce} task. The reducers perform their joins
by dividing the space that corresponds to them into stripes and
performing plane sweep for each tile. Duplicate results are avoided by
reporting a join pair only at the tile with the smallest id where the
two objects commonly appear. 
In the {\em Hadoop-GIS} system \cite{AjiWVLL0S13} spatial joins are
computed in a similar fashion. The space is divided by a grid to tiles
and the objects in each tile are stored locally at nodes in the
HDFS. 
A {\em global index} that is shared between nodes is used to find
the HDFS files where the data of each tile are located. The local data at each
node are also locally indexed. Local joins are performed by each node
separately and the results are merged.
The {\em SpatialHadoop} system \cite{EldawyM15} follows a similar approach
where a global index per dataset is stored at a Master node, however, 
different datasets may have different partitioning that could
pre-exist before queries. 
If the two join inputs are partitioned differently, then two options exist
(1) use existing partitions and perform joins for every pair of
partitions that overlap (they could be many) or (2) re-partition the
smaller file using the same partition boundaries as the larger
file and apply MASJ. The cost of each the two options is estimated and the cheapest
one is selected accordingly. 
A query optimizer for MapReduce-based spatial join algorithms
is presented in \cite{SabekM17}.

Implementations of spatial joins using
Spark, where the data, indexes, and intermediate results are shared in
the memories of all nodes in a cluster have also been proposed
\cite{YouZG15,YuWS15,000100LCZG16}, with a focus on effective spatial indexing of
the {\em resilient distributed datasets} (RDDs) that are generated during
the process. 
Processing spatial joins in parallel with a focus on
minimizing the cost of the refinement step
was studied by Ray et al. \cite{RaySBJ14},
During
data partitioning, the exact geometries of objects are clipped and
distributed to partitions (that may be handled by different nodes) in
order to avoid any communication between nodes when object pairs
whose MBRs intersect are refined.
Pandey et al. \cite{PandeyKNK18} conducted an experimental 
evaluation between 
parallel and distributed spatial data management systems
where, among other queries, spatial joins were tested.
A recent piece of related work \cite{Kipf18} studies the join
between streaming points and static polygons focusing on accelerating the refinement step using modern hardware.

%\section{Boosting in-Memory Spatial Joins}\label{sec:tuning}
\section{Tuning PBSM}\label{sec:tuning}

As discussed in the previous section, the most popular spatial join
framework follows the multi-assignment, single-join (MASJ)
paradigm of PBSM.
The reasons behind this can be summarized as follows:
\begin{itemize}
\item{PBSM does not assume any preprocessing or indexing of the data
    before the join, hence it can be applied on dynamically generated
    spatial data.}
\item{The
partitions define independent join tasks that can easily be
distributed and/or parallelized.}
\item{The number of join tasks is the same as the number of partitions (as
opposed to the number of join tasks of SAMJ approaches which can be
much higher).}
\item{Producing duplicate results can be easily avoided.}
\item{Implementing this approach is fairly easy.}
\item{Previous studies \cite{NobariQJ17} have shown that the 
performance of PBSM can hardly be beaten by
more sophisticated approaches based on indexing or adaptive
partitioning.}
\end{itemize}
In this section, 
we explore the directions along which we can tune PBSM in order to
improve its performance.
These include determining the number and type of partitions (tiles or stripes), considering
alternative duplicate avoidance mechanisms, and choosing the axis along
which we perform plane sweep in each partition.
We also study the flexibility that these directions give to
balancing the computation load in parallel spatial
join evaluation. 
In Section \ref{sec:exp}, we evaluate the effect that all these
parameters have in the performance of the algorithm when joining
big spatial datasets.

\subsection{One-dimensional Partitioning}\label{sec:onedim}
\begin{figure}[htb]
\centering
  \includegraphics[width=0.45\columnwidth]{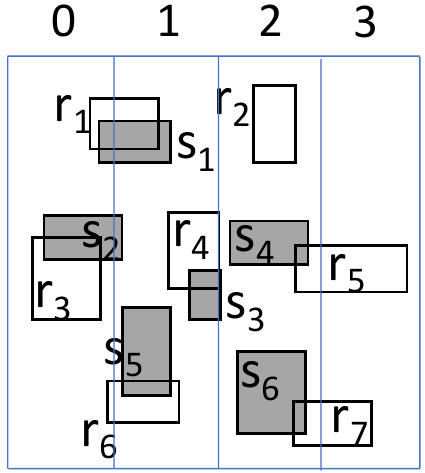}
  \caption{Example of 1D partitioning}
  \label{fig:stripes}
\end{figure}

The default partitioning approach followed by PBSM is by a 2D
grid, as shown in Figure \ref{fig:pbsm}.
Still, the same algorithm can be applied if we partition the data
space in 1D {\em stripes}, as shown in Figure \ref{fig:stripes}.
The stripes can be horizontal or vertical.
Such a partitioning has already been considered by an
{\em external memory plane sweep} join algorithm \cite{ArgePRSV98};
however, the objective of the partitioning there was to define the
stripes in a way such that the ``horizon'' of the sweep line (which
runs along the axis of the stripes) fits in memory.
That is, the goal of partitioning was that the {\em active} sets
of rectangles during a stripe-to-stripe plane
sweep join is small enough to fit in memory.
Since in this paper, we deal with in-memory joins, we do not
consider this factor, but we study how the
number of partitions affects the computational cost of the spatial join
(as we do for the 2D partitioning case).

%\subsection{Handling Duplicates}\label{sec:avoidance}
%In what follows, we discuss two approaches for handling duplicates.

\subsection{Duplicate Elimination}
Dittrich and Seeger \cite{DittrichS00} presented a simple but
effective approach for eliminating duplicate results in PBSM.
A rectangle pair is reported by a partition-partition join only if the top-left
corner of their intersection area is inside the spatial extent of the
partition.
For example, consider the join of Figure \ref{fig:pbsm}.
The pair of intersecting rectangles $(r_1,s_1)$ can be found in both
tiles (0,0) and (0,1). However, the result will only be reported in
tile (0,0), which contains the top-left corner of the intersection. In
other words, the join result will be computed in tile (0,1) but not
reported. Hence, for each rectangle pair found to intersect, a
{\em duplicate} test is performed. Let $[r.x_l,r.x_u]$ and
$[r.y_l,r.y_u]$ be the projections of rectangle $r$ on the $x$ and
$y$ axis, respectively. Let $[T.x_l,T.x_u]$ and $[T.y_l,T.y_u]$ be
the corresponding projections of a tile.
The duplicate test for pair $(r,s)$, found to intersect in tile $T$,
is the condition:
\begin{equation}\label{eq:ditt}
\max\{r.x_l,s.x_l\}\ge T.x_l \land \max\{r.y_l,s.y_l\}\ge T.y_l
\end{equation}
%We call this duplicate avoidance approach {\em Generate and Duplicate
%  Test} (GDT).

\vspace{1ex}\noindent{\bf Application to 1D partitioning.}
For the case of 1D partitioning, the duplicate test needs to apply a single
comparison (as opposed to the two comparisons of Eq. \ref{eq:ditt}).
For example, if the stripes are vertical (as in Figure
\ref{fig:stripes}), a join result is reported only if
$\max\{r.x_l,s.x_l\}\ge T.x_l$.

\subsection{Choosing the Sweeping Axis}\label{sec:sweepaxis}
When applying plane sweep for a tile (or stripe) $T$, we have to decide along
which axis we will sort the rectangles and then sweep them.
As we will show in Section \ref{sec:exp} choosing the proper axis can
make a difference. For this purpose, we devise a model which, given
the sets of rectangles $R_T$, $S_T$ that are assigned to tile $T$, 
determines the sweeping axis to be used. The key idea is
to estimate, for each axis, how many pairs of rectangles 
from $R_T\times S_T$ intersect along this axis.
For example, if the sweeping axis is $x$, 
according to Algorithm~\ref{algo:fs}, rectangle pairs that 
$x$-intersect,
are essentially \emph{candidate} or \emph{intermediate} results; the
algorithm must then verify in Line~8 or 15 if their $y$-projections
also intersect.
Hence, choosing as sweeping axis the one which produces the  smallest
number of candidates pairs, can reduce the cost of plane sweep.

To estimate the number of intersecting projections per axis, we
compute histogram statistics.
In specific, we sub-divide the $x$ and $y$
projections of the tile $T$ into a predefined number of partitions $k$.
Then, we count how many rectangles from $R$ and how many from $S$,
$x$-intersect each $x$-division of the tile;
the procedure for $y$ partitions is symmetric.
In this manner, we construct four histograms $H_R^x$, $H_R^y$,
$H_S^x$, $H_S^y$ of $k$ buckets each.
The number $I^x_T$ of rectangles in $R_T\times S_T$  that $x$-intersect can
then be approximated
by accumulating the product of the corresponding histogram buckets,
i.e.,
\begin{equation}\label{eq:histo}
I^x_T=\sum_{i=0}^{k}\{H_R^x[i]\cdot H_S^x[i]\}
\end{equation}
 The smallest of $I^x_T$ and $I^y_T$ determine the chosen sweeping axis
 (i.e., $x$ or $y$).
For large tiles (compared to the size of the rectangles), we set
$k=1000$,
while for small tiles $k$ is the number of times the tile's extent is
larger than the average extent of the rectangles. 
In practice, it would be expensive to use all rectangles in $T$ in the
histogram construction. So, we use a sample of rectangles from $R_T$
and $S_T$ for this purpose. Specifically, for every $\phi$ rectangles
that are assigned to tile $T$, we use one for histogram construction.
We set $\phi=100$ by default because it can produce good enough
estimates at a low overhead. 
%\nikos{why do you need this? I thought you can do the histograms
%  etc. at the time of partitioning. That is, for a sample of the
%  hashed rectangles per partition, you can compute the histogram
%  statistics.}

\vspace{1ex}\noindent{\bf Application to 1D partitioning.} Our model
can be straightforwardly applied in case of a 1D dimensional
partitioning; histogram statistics are now computed for the contents
of the vertical or horizontal stripes and the entire domain on the
other dimension. However, our tests in Section~\ref{sec:exp} showed that 
in practice, if the partitioning axis is $x$, the best sweeping axis
is always $y$ and vice versa. 
%there is a more effective way to determine the sweeping axis based on the partitioning dimension.
%\fix{Intuitively, we will have to choose the axis that.....We now propose a simple test that ....}

\subsection{Parallel Processing}\label{sec:parallel}
We parallelize evaluation by splitting each of the partitioning and joining
phases of the algorithm into parallel and independent tasks, while
trying to minimize the synchronization requirements between the
threads. While the parallel algorithm that we outline here is designed
for a single, multi-core machine, it can also be applied (with minor
changes) to a cluster of machines.
The steps for parallelizing the spatial join to $m$ threads are as follows:
\begin{enumerate}
\vspace{0.5ex}
{\setlength\itemindent{-25pt} \item[]{\bf Partitioning phase}}
\item Determine a division of each input $R$ and $S$ into $m$
  equi-sized parts arbitrarily.
\item Initiate $m$ threads. Thread $i$ reads the $i$-th part of input
  $R$ and {\em counts} how many rectangles should be assigned to each
  of the space partitions (tiles or stripes). Thread $i$ repeats the
  same process for the $i$-th part of input $S$. Let $|R_T^i|, |S_T^i|$
  be the numbers of rectangles counted by thread $i$ for tile $T$ and
  $R$, $S$, respectively. \label{item:counting}
\item Compute $|R_T|=\sum_i^m|R_T^i|$ and $|S_T|=\sum_i^m|S_T^i|$ for each tile
  $T$. Allocate two memory segments for $|R_T|$ and $|S_T|$ rectangles
  of each partition $T$. 
\item Initiate $m$ threads. Thread $i$ reads the $i$-th parts of inputs
  $R$ and $S$ and partitions them.
  The memory allocated for each of $|R_T|$ and $|S_T|$ is logically
  divided into $m$ segments based on the $|R_T^i|$'s and $|S_T^i|$'s.
  Hence, thread $1$ will
  write to the first $|R_T^1|$ positions of $|R_T|$, thread $2$ to the
  next $|R_T^2|$ positions, etc.
  After all threads complete partitioning, we will have the entire
  set of rectangles that fall in each tile continuously in memory. \label{item:partitioning}
 \vspace{0.5ex}
{\setlength\itemindent{-25pt} \item[]{\bf Joining phase}}
\item Construct two sorting tasks for each tile $T$ (one for $R_T$ and
  one for $S_T$). Assign the sorting tasks to the $m$ threads. 
\item Construct a join task for each tile $T$ (one for $R_T$ and
  one for $S_T$). Assign the join tasks to the $m$ threads. 
\end{enumerate}
\vspace{1ex}

\noindent Step \ref{item:counting} is applied in order to make proper memory allocation and
prevent expensive dynamic allocations.
It also facilitates the output of parallel partitioning for each tile $T$
to be continuous in memory during step \ref{item:partitioning}. 
When the model presented in Section \ref{sec:sweepaxis} is used, the
histograms are computed when the input data are read (i.e., in either of
steps  \ref{item:counting} and \ref{item:partitioning}).
%If the GBDA approach is used for handling duplicate, the memory
%allocated for each tile is further divided into sub-segments for
%classes $A$-$D$ (or $A$-$B$ in 1D partitioning). In addition, the
%number of join tasks increases by a factor of nine, which leaves room
%for better load balancing.

%\fix{In a parallel processing environment, the number and cost
%  variance in the join tasks can affect load balancing.....}
%\pbour{We should talk about the mini-joins effect}

\eat{
\subsection{Factors Affecting Performance}\label{sec:factors}
\nikos{This section will be removed}
Summing up, the factors that may affect the performance of a PBSM-like
spatial join are the following:
\begin{itemize}
\item{1D or 2D partitioning?}
\item{how many partitions?}
\item{along which direction do we sweep in each partition?}
\item{which duplicate avoidance method to choose?'}
\item{how to achieve load balancing in parallel evaluation?}
\end{itemize}
}

\section{Experimental Analysis}\label{sec:exp}
In this section, we test the effect of the different parameters in the
design of a PBSM-like spatial join. We first describe the experimental
setup and then perform experiments, which test how each factor affects
the performance of the algorithm and come up with empirical rules for
selecting parameter values.

\subsection{Setup}\label{sec:exp:setup}
We experimented with a number of publicly available datasets
\cite{EldawyM15}.\footnote{http://spatialhadoop.cs.umn.edu/datasets.html}
For each dataset, we computed the MBRs of the objects and came up
with a corresponding collection of rectangles.
The datasets are normalized so that the coordinates in each dimension
take values in $[0,1]$. 
Table \ref{tab:datasets} provides statistics about the datasets that we
used.
%\nikos{normalize the statistics for T4 and O3} 
The first three datasets are from the collection Tiger 2015 and
the last three from the collection OpenStreetMap (OSM).
Next to each dataset name we put a short alias indicating its
order in the Tiger or OSM collection (i.e., $O3$ means the 3rd dataset
from OSM).
The cardinalities of
the datasets range from $2.3$M to $115$M objects
and we tested joins
having inputs from the same collection, having similar or various
scales.
The last two columns of the tables are the 
relative (over the entire space) average length
of the rectangle projections at each axis.

\eat{
\begin{table}[!t]
\centering
\caption{Real datasets used in experiments}
\label{tab:datasets}
%\vspace{-2ex}
\small
\begin{tabular}{|c|c|c|c|c|c|c|}\hline
\textbf{dataset} &\textbf{alias} &\textbf{cardinality} & \textbf{avg. $x$-extent}&\textbf{avg. $y$-extent}\\\hline\hline
AREAWATER &T2                 &$2.3$M		&$0.00259606$ 	&$0.00197538$ \\
LINEARWATER &T5               &$5.8$M		&$0.00795765$ 	&$0.00627023$ \\
ROADS &T8                &$20$M		&$0.0044686$ 	&$0.00348621$ \\
Lakes &O5                &$8.4$M		&$0.00756612$ 	&$0.00490276$ \\
Parks &O6                &$10$M		&$0.00594295$ 	&$0.00384502$ \\
Roads &O9                &$72$M		&$0.00379772$ 	&$0.00280823$ \\
\hline
\end{tabular}
\end{table}
}

\begin{table}[!t]
\centering
\caption{Real datasets used in experiments}
\label{tab:datasets}
%\vspace{-2ex}
\small
\begin{tabular}{|c|c|c|c|c|c|c|}\hline
\textbf{dataset} &\textbf{alias} &\textbf{cardinality} & \textbf{avg. $x$-extent}&\textbf{avg. $y$-extent}\\\hline\hline
AREAWATER &$T2$                 &$2.3$M		&$0.000007230$ 	&$0.000022958$ \\
  EDGES &$T4$ & $70$M & $0.000006103$ &$0.00001982$ \\
  LINEARWATER &$T5$               &$5.8$M		&$0.000022243$ 	&$0.000073195$ \\
ROADS &$T8$                &$20$M		&$0.000012538$ 	&$0.000040672$ \\
Buildings &$O3$ & 115M &  $0.00000056$ & $0.000000782$ \\
Lakes &$O5$                &$8.4$M		&$0.000021017$ 	&$0.000028236$ \\
Parks &$O6$                &$10$M		&$0.000016544$ 	&$0.000022294$ \\
Roads &$O9$                &$72$M		&$0.000010549$ 	&$0.000016281$ \\
\hline
\end{tabular}
\end{table}

We implemented the spatial join algorithm (all different versions) in
  C++ and compiled it using \texttt{gcc} (v4.8.5) with flags \texttt{-O3}, \texttt{-mavx} and \texttt{-march=native}. For multi-threading, we used OpenMP. All experiments were run on a machine with 384 GBs of RAM and a dual 10-core Intel(R) Xeon(R) CPU E5-2630 v4 clocked at 2.20GHz running CentOS Linux 7.3.1611; with hyper-threading, we were able to run up to 40 threads.
The reported runtimes include the costs of partitioning both
datasets and then joining them.
%\nikos{will we have an experiment with a cost-breakdown?}

\subsection{Selecting Sweeping Axis}
In the first experiment, we test the effect that the selection of the
axis we sweep along (either $x$ or $y$) has on the performance of the
algorithm.
For this purpose, we chose not to partition the data, but ran a
single-threaded plane-sweep join using Algorithm~\ref{algo:fs} in the
entire $[0,1]\times[0,1]$ dataspace (i.e., modeling the case of a single tile).
Table \ref{tab:sweepaxis} reports the execution times per query. We observe that sweeping along the wrong axis may even double the cost of spatial join.
%In \fix{all}\note{need more queries} cases, sweeping along the
%$x$ axis is much cheaper (costs about half) compared to sweeping along $y$. 
The last column of the table reports the details of running %sweeping axis recommended by
our model (Eq. \ref{eq:histo}).
%of Section~\ref{sec:sweepaxis}.
Our model was able to accurately determine the proper sweeping axis in
all cases.
Note that the cost of this decision-making process is
negligible compared to the partitioning and joining cost; even for the
largest queries, our model needs less than 10
milliseconds.
%\nikos{do you do sampling here, or do you use all
%  rectangles to build the histograms. Also, mention the value of $k$
%  (if it is fixed, you can mention it earlier or in section 3.3.)}
%\fix{By looking at the statistics of the data... (explain result)}

\begin{table}[!t]
\centering
\caption{Sweeping axis effect; queries ordered by runtime}
\label{tab:sweepaxis}
%\vspace{-2ex}
\small
\begin{tabular}{|c|c|c|c|c|}\hline
\multirow{2}{*}{\textbf{query}} &\multicolumn{2}{c|}{\textbf{sweeping axis}} &\multicolumn{2}{c|}{\textbf{adaptive model}} \\\cline{2-3}\cline{4-5}
% &$x$ &$y$ &$\sum_{i=0}^{k}\{H_R^x[i]\cdot H_S^x[i]\}$ &$\sum_{i=0}^{k}\{H_R^y[i]\cdot H_S^y[i]\}$\\\hline\hline
 &$x$ &$y$ &$I^x$
      &$I^y$\\\hline\hline
$T2 \bowtie T5$ &$8.94$s		&$16.96$s &$8,\!376$ &$19,\!232$\\
$T2 \bowtie T8$ &$24.52$s	& $40.72$s &$8,\!895$ &$18,\!660$\\
$O5 \bowtie O6$             &$24.92$s	&$66.06$s &$2,\!692$ &$12,\!279$\\
$O6 \bowtie O9$          &$216.88$s &$444.19$s &$3,\!989$ &$11,\!510$\\
$T4 \bowtie T8$ &$674.50$s &$1,\!360.92$s &$8,\!135$  &$19,\!406$\\
$O9 \bowtie O3$ &$926.14$s &$1,\!681.30$s&$4,\!535$ &$11,\!529$\\
%\pbour{we need more queries here...}\\
\hline
\end{tabular}
\end{table}

\begin{figure*}[!ht]
\begin{center}
\fbox{
{\small $xx$}
%\hspace{0.5ex}
\includegraphics[width=0.08\columnwidth]{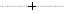}
\hspace{1ex}
{\small $yy$}
%\hspace{0.5ex}
\includegraphics[width=0.08\columnwidth]{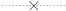}
\hspace{1ex}
{\small $xy$}
%\hspace{0.5ex}
\includegraphics[width=0.08\columnwidth]{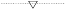}
\hspace{1ex}
{\small $yx$}
%\hspace{0.5ex}
\includegraphics[width=0.08\columnwidth]{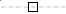}
}
\end{center}
\begin{tabular}{cccc}
\hspace{-2ex}\includegraphics[width=0.245\linewidth]{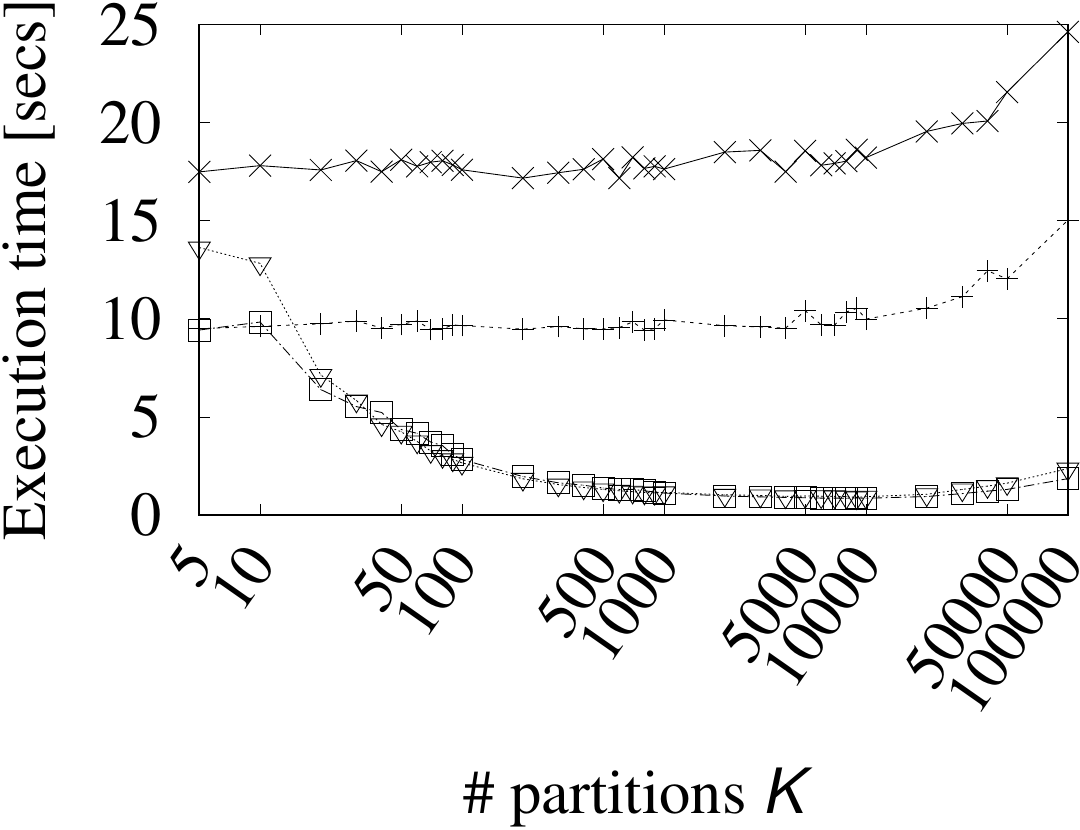}
&\hspace{-0.5ex}\includegraphics[width=0.245\linewidth]{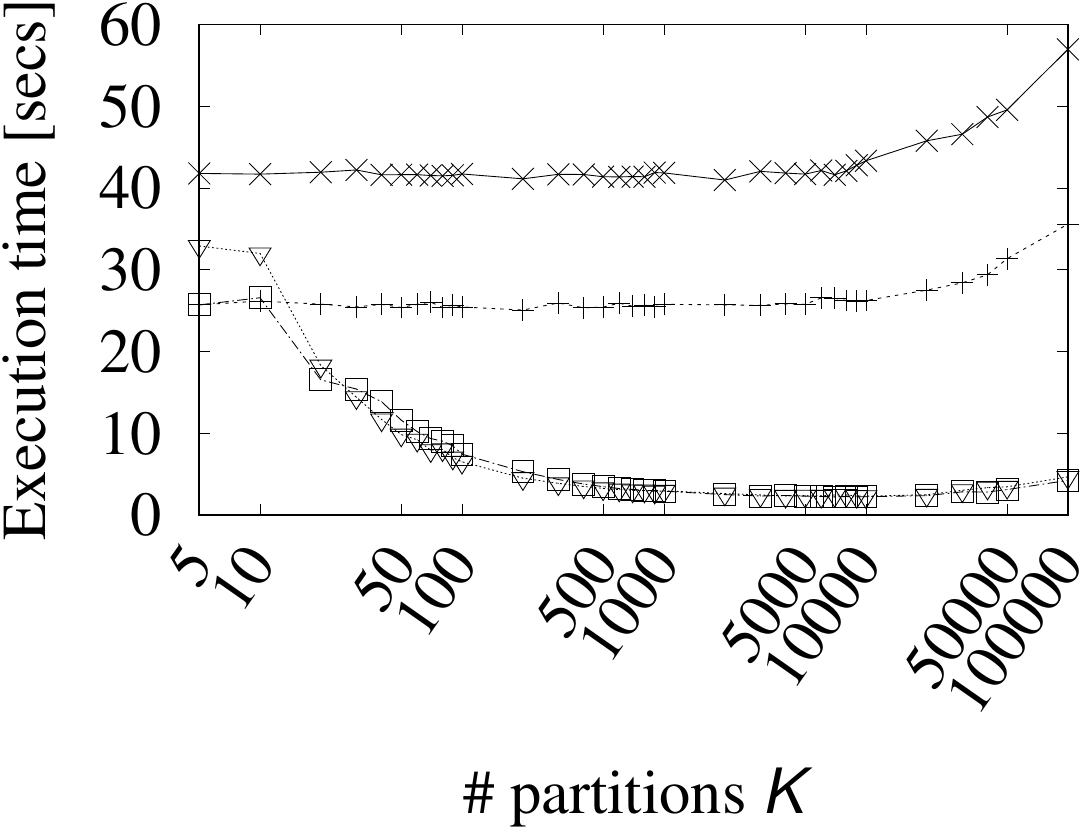}
&\hspace{-0.5ex}\includegraphics[width=0.245\linewidth]{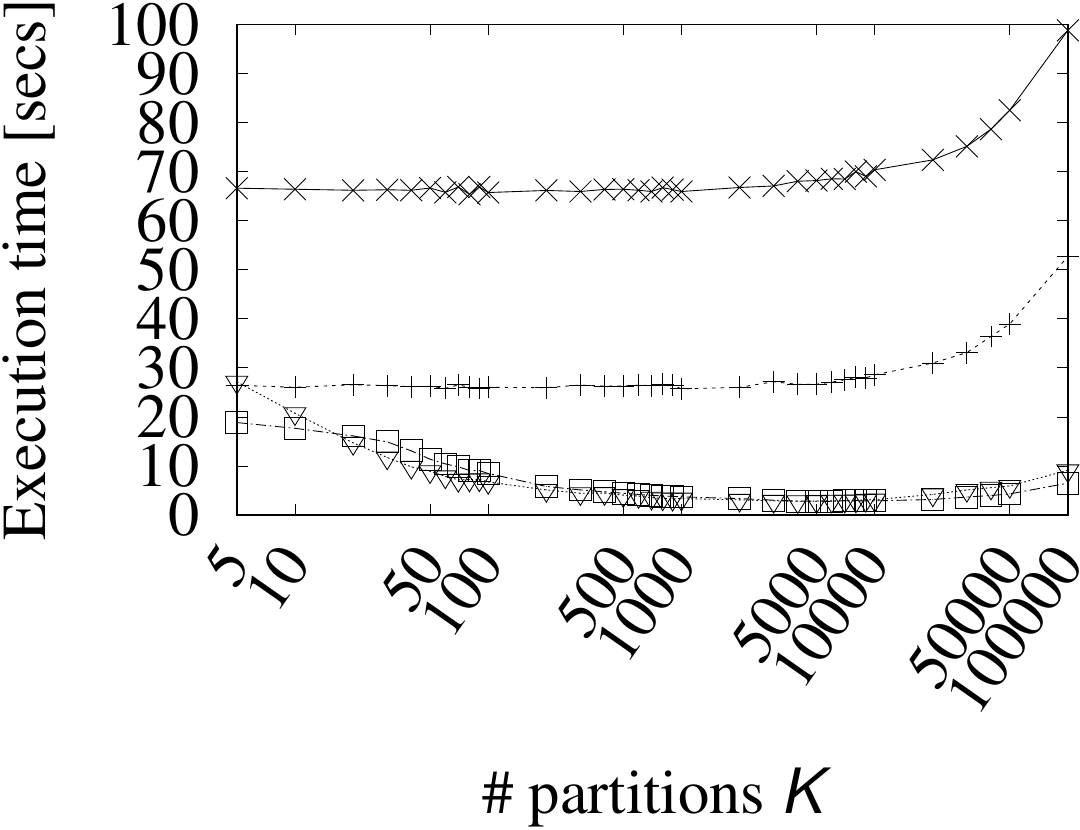}
&\hspace{-0.5ex}\includegraphics[width=0.245\linewidth]{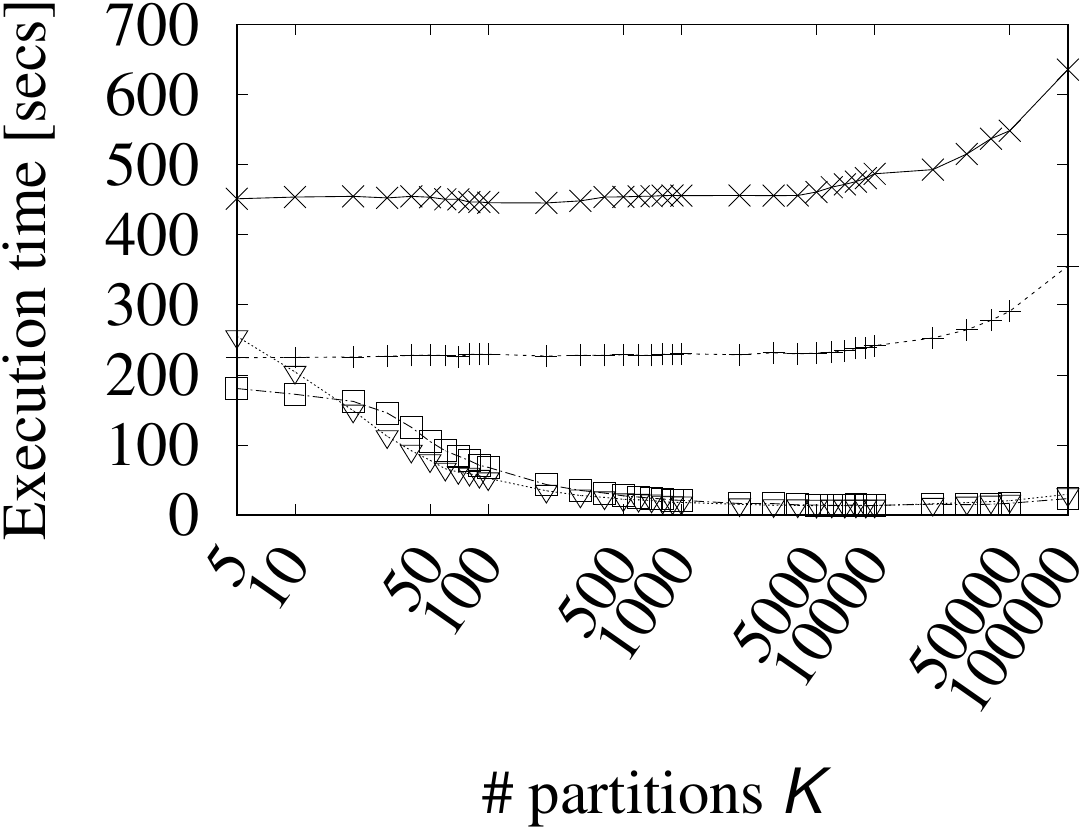}\\
(a) $T2 \bowtie T5$ &(b) $T2 \bowtie T8$ &(c) $O5 \bowtie O6$ &(d) $O6 \bowtie O9$\\
\end{tabular}
%\vspace*{-2ex}
\caption{Tuning 1D partitioning: total execution time}
\label{fig:tuning_1d}
%\vspace{-5ex}
\end{figure*}
\begin{figure*}[!ht]
\begin{center}
\fbox{
{\small partitioning}
%\hspace{0.5ex}
\includegraphics[width=0.08\columnwidth]{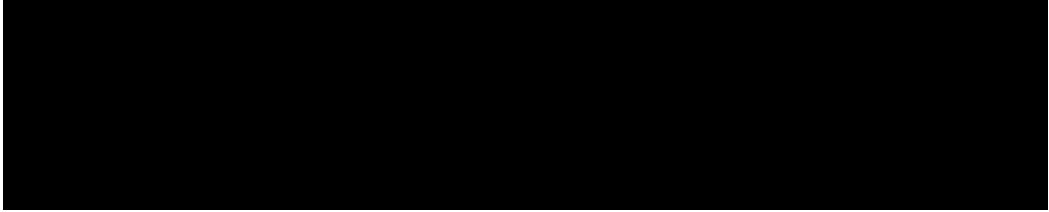}
\hspace{1ex}
{\small joining}
%\hspace{0.5ex}
\includegraphics[width=0.08\columnwidth]{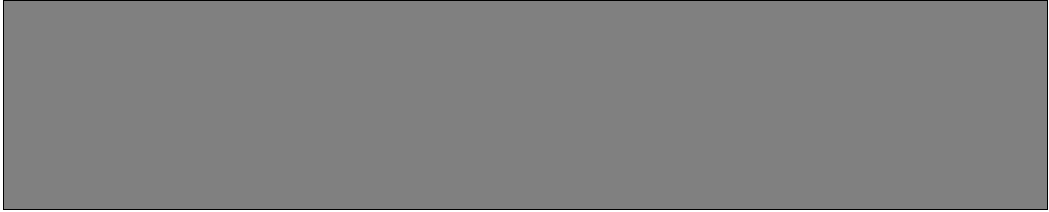}
}
\end{center}
\begin{tabular}{cccc}
\hspace{-2ex}\includegraphics[width=0.245\linewidth]{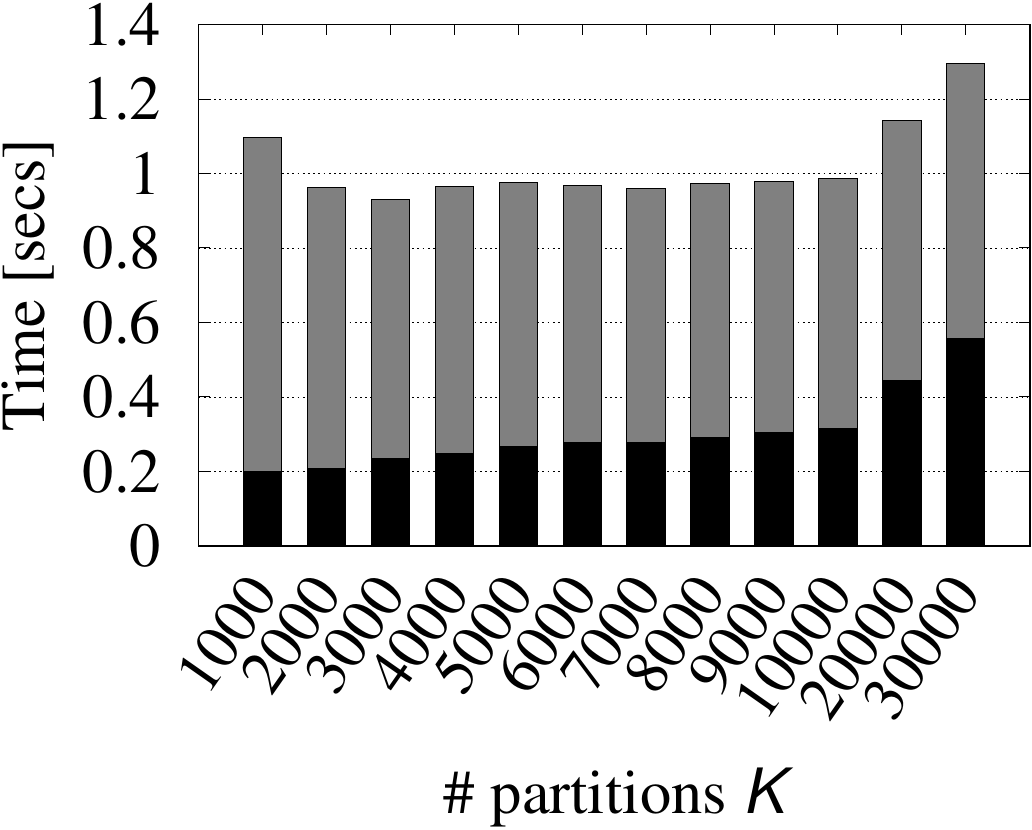}
&\hspace{-1ex}\includegraphics[width=0.245\linewidth]{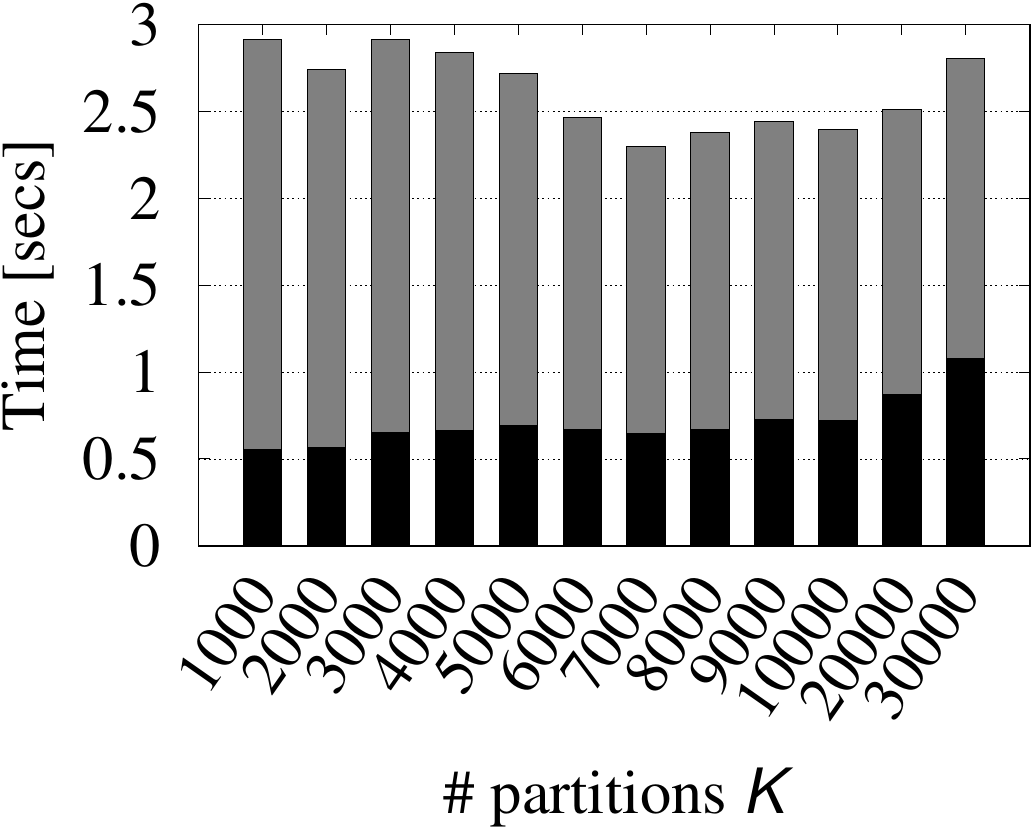}
&\hspace{-1ex}\includegraphics[width=0.245\linewidth]{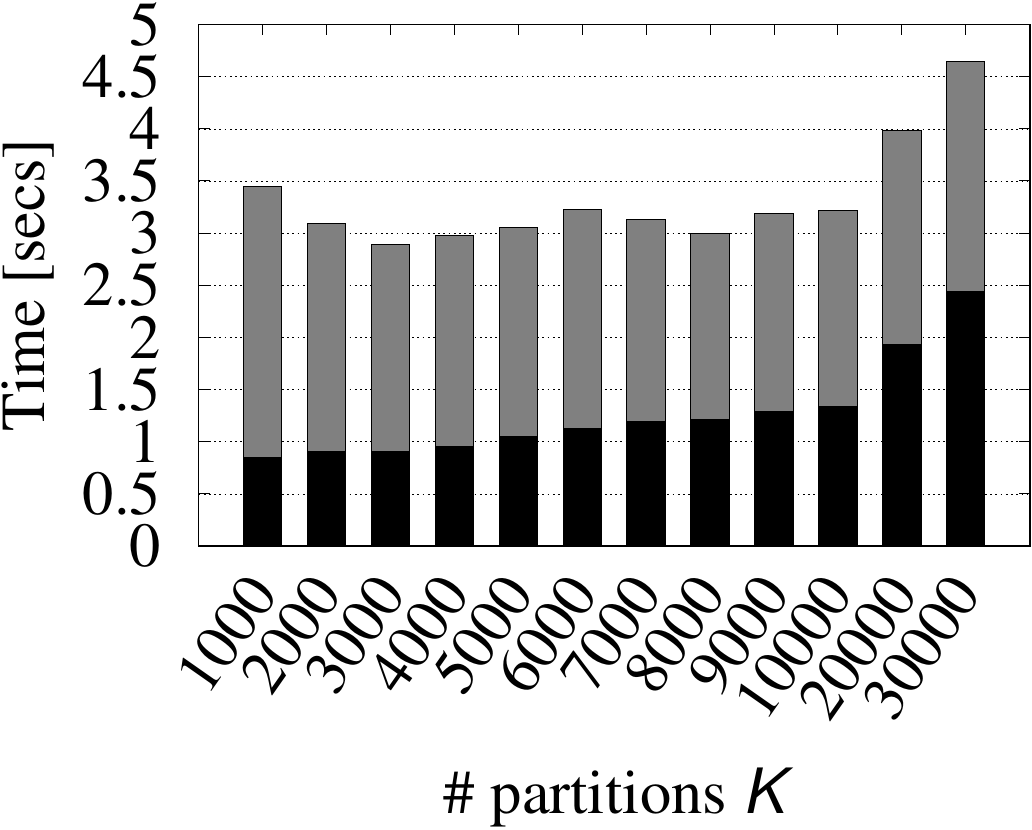}
&\hspace{-1ex}\includegraphics[width=0.245\linewidth]{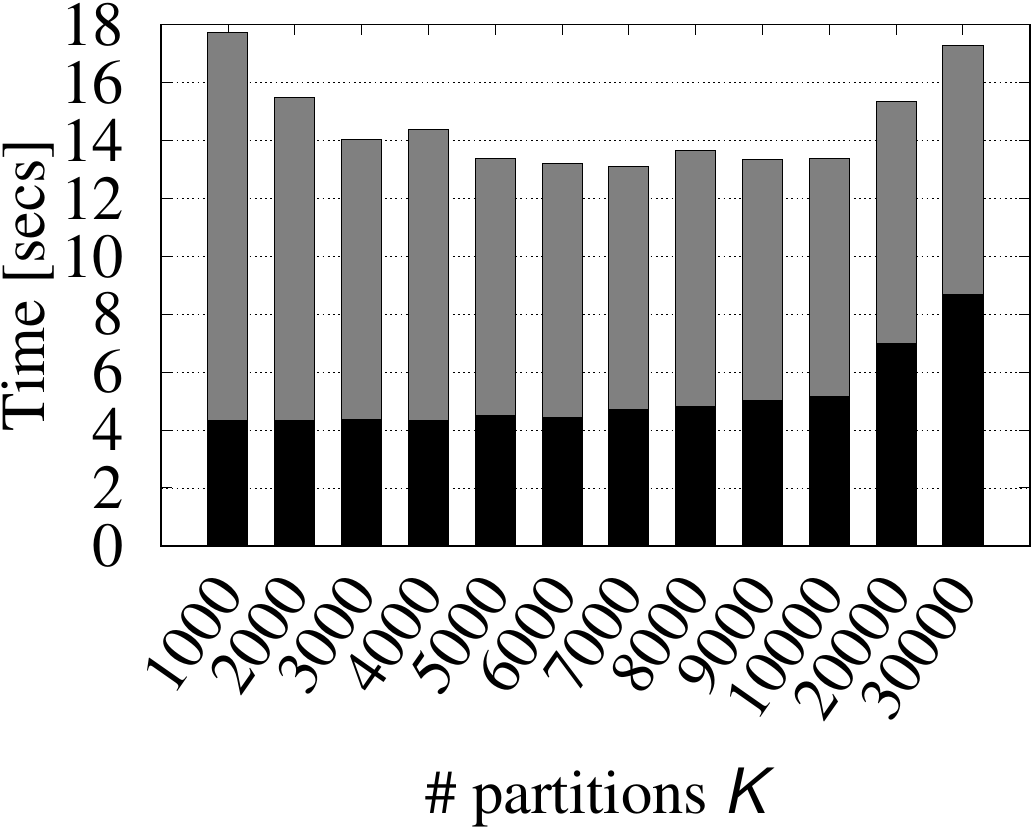}\\
%\scriptsize{\# partitions} &\scriptsize{\# partitions} &\scriptsize{\# partitions} &\scriptsize{\# partitions}\\
(a) $T2 \bowtie T5$ &(b) $T2 \bowtie T8$ &(c) $O5 \bowtie O6$ &(d) $O6 \bowtie O9$\\
\end{tabular}
%\vspace*{-2ex}
\caption{Tuning 1D partitioning: time breakdown}
\label{fig:tuning_1d_breakdown}
%\vspace{-1ex}
\end{figure*}

\subsection{Evaluation of Partitioning}
In the next set of experiments, we investigate the impact of
partitioning to the performance of the algorithm.
We first tune 1D and 2D-based PBSM and then compare the two partitioning approaches to each other.
\subsubsection{Tuning 1D Partitioning}

%In the next experiment, we evaluate the performance of the algorithm
%when 1D partitioning is used.
We varied the number $K$ of (uniform) 1D partitions and for each such number,
we report the cost of the algorithm for four spatial join queries in
Figure~\ref{fig:tuning_1d}.
We tested all combinations of partitioning and sweeping axes.
%; \fix{note that combinations which use our adaptive model to select
%the sweeping axis were found inefficient and hence omitted to keep
%plots as clear as possible}.
For
example, $xy$ denotes partitioning along the $x$ axis (to vertical
stripes) and sweeping along the $y$ axis. 
%For all queries, we use the GDT approach for handling duplicates.
We can make the following observations from the plots. First, and
foremost, if the sweeping axis is the same as the partitioning axis
(i.e., cases $xx$ and $yy$), the join cost does not drop when we
increase the number of partitions $K$. This is expected because,
regardless the number of partitions, case $xx$ or $yy$ is
equivalent to having no partitions at all and sweeping along the $x$
or $y$ axis in the entire space. When $K$  is too
large, the costs of  $xx$ and $yy$ increase because the partitions
become very narrow and replication becomes excessive.
The second observation is that the performance of cases $xy$ and $yx$
improves with $K$  and, after some point, i.e., $K=2,\!000$, they
converge to the same (very low) cost. The costs of both $xy$ and $yx$
starts to increase again when $K>10,\!000$, at which point we start
having significant replication (observe the average $x$- and
$y$-extent statistics in Table \ref{tab:datasets}).
%Between $xy$ and $yx$, $xy$ performs better, especially when the
%number of partitions is small $K<2000$ because 
%., we partition along the $x$ axis to vertical stripes and then 
%number of partitions is small
%\fix{++}
Figure \ref{fig:tuning_1d_breakdown} breaks down the total cost to partitioning and
joining for the $xy$ case in the range of $K$ values where the best performance was witnessed. The cost of sorting the partition is included in the join
cost. As expected, as $K$ increases the cost of partitioning increases
and the cost of the join drops, up to a point (around $K=10,\!000$) after
which the cost of partitioning significantly increases without
providing any improvement in the performance of the join.
Overall, the $xy$ case is marginally better than the $yx$
  case and the lowest runtime is achieved when the
$x$-extent of the partitions (i.e., the narrow side of the stripes) is
about 10 times larger than the average $x$-extent of the
rectangles. In this case, the chances that a rectangle is
replicated to neighboring stripes are small and at the same time, the stripes are
narrow enough for plane sweep to be effective (i.e., the chance that
a candidate pair that $x$-intersects also $y$-intersects is not low).
For the rest of our analysis we use $xy$ as the default setup for 1D partitioning.

%\fix{report number of empty partitions}

%\fix{report any correlation of best partitioning to average rectangle
% extent (vs. $x$-extent of each partition)} 

\begin{figure*}[!ht]
\begin{center}
\fbox{
{\small $x$}
%\hspace{0.5ex}
\includegraphics[width=0.08\columnwidth]{figures/plus.pdf}
\hspace{1ex}
{\small $y$}
%\hspace{0.5ex}
\includegraphics[width=0.08\columnwidth]{figures/x.pdf}
\hspace{1ex}
{\small adaptive model}
%\hspace{0.5ex}
\includegraphics[width=0.08\columnwidth]{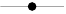}
}
\end{center}
\begin{tabular}{cccc}
\hspace{-2ex}\includegraphics[width=0.245\linewidth]{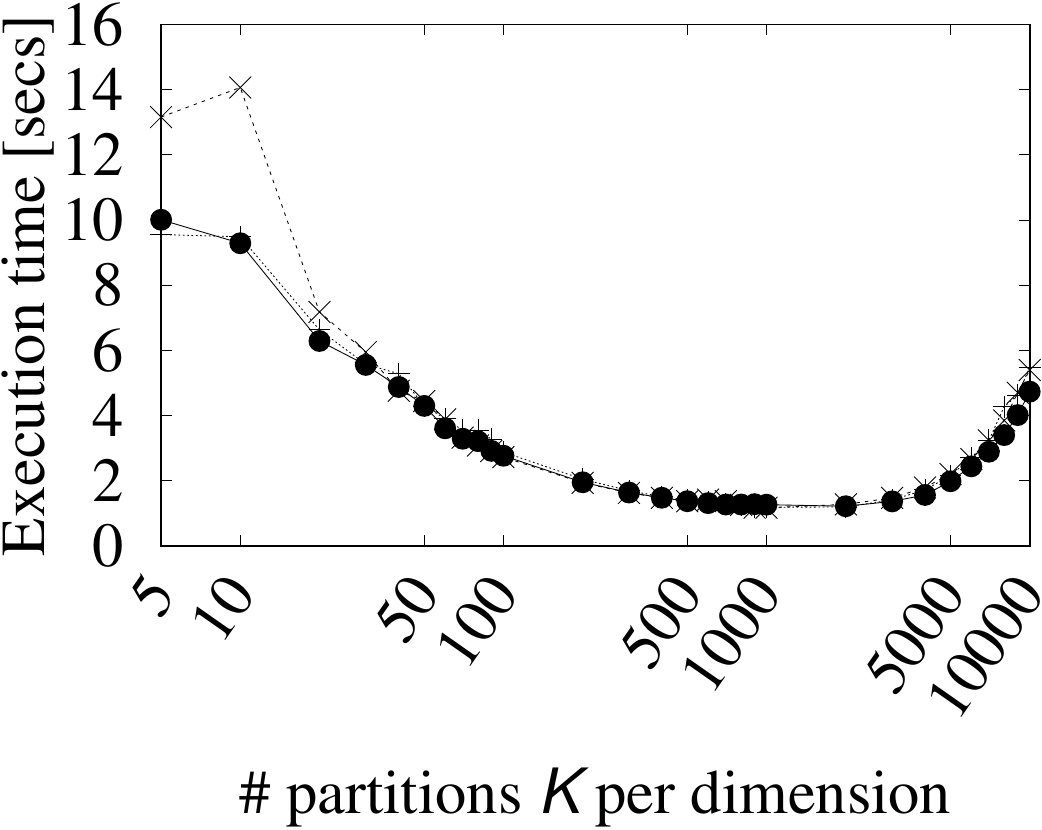}
&\hspace{-1ex}\includegraphics[width=0.245\linewidth]{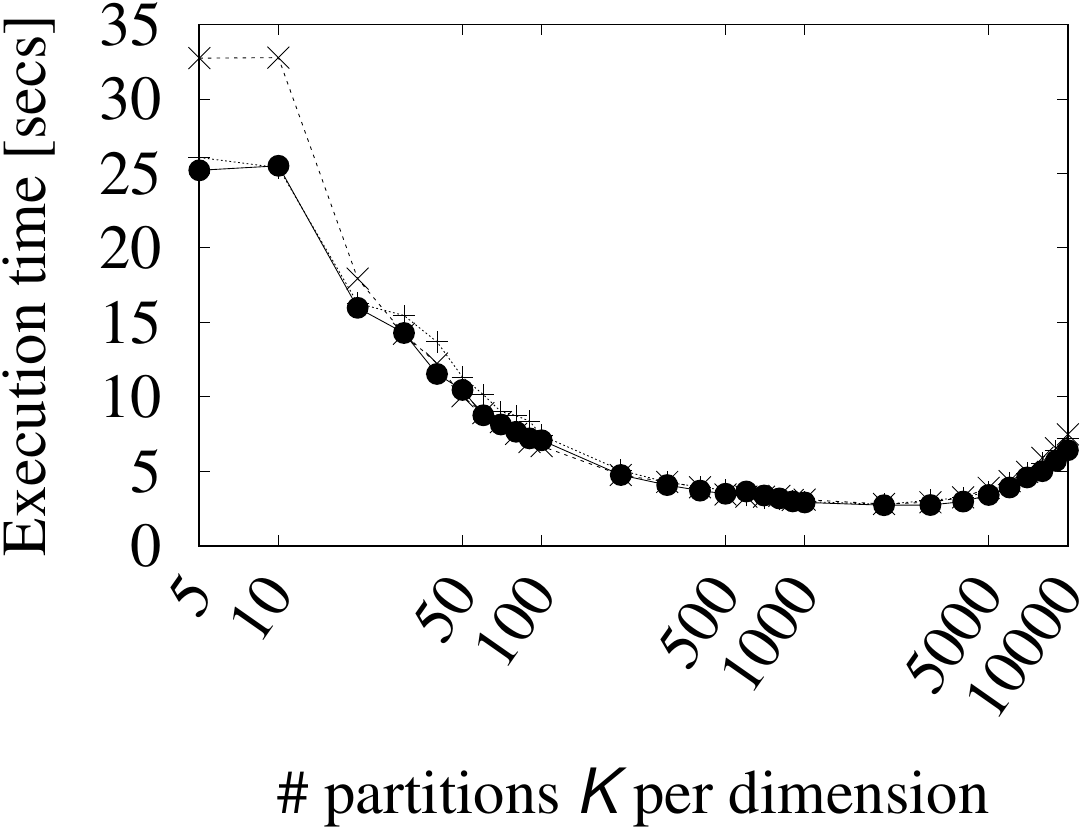}
&\hspace{-1ex}\includegraphics[width=0.245\linewidth]{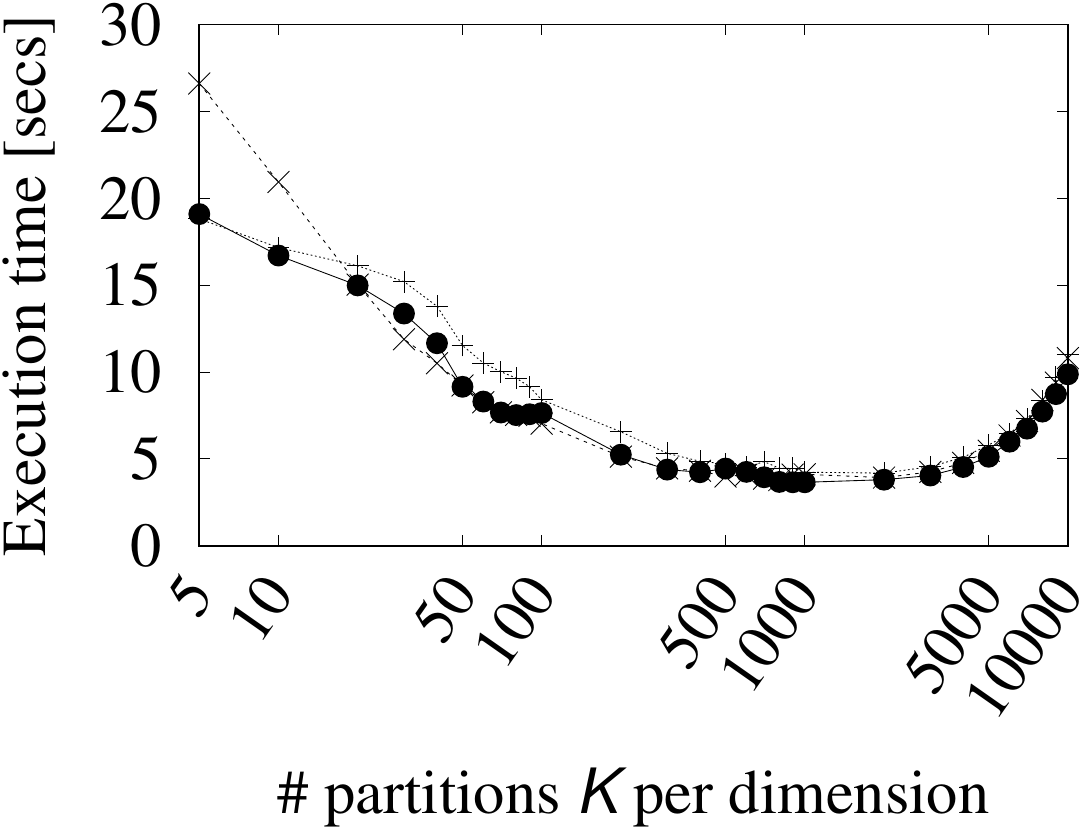}
&\hspace{-1ex}\includegraphics[width=0.245\linewidth]{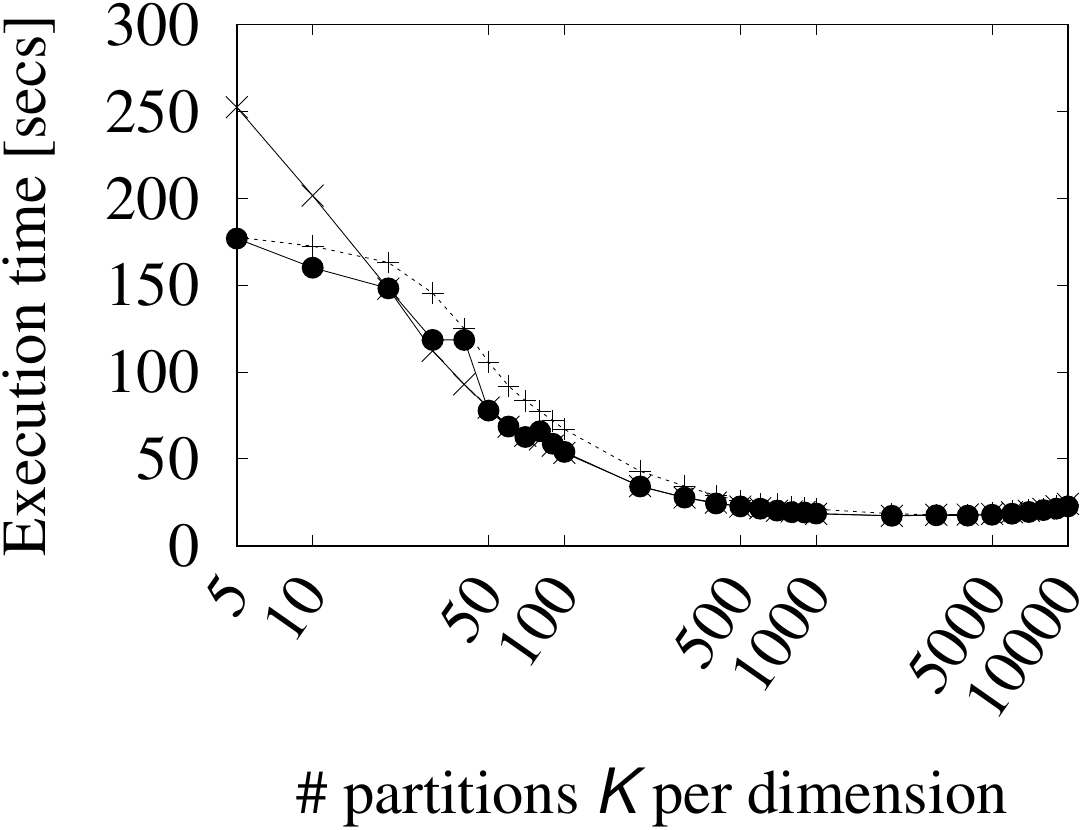}\\
$T2 \bowtie T5$ &$T2 \bowtie T8$ &$O5 \bowtie O6$ &$O6 \bowtie O9$\\
\end{tabular}
%\vspace*{-2ex}
\caption{Tuning 2D partitioning: total execution time}
\label{fig:tuning_2d}
%\vspace{-1ex}
\end{figure*}
\begin{figure*}[!ht]
\begin{center}
\fbox{
{\small partitioning}
%\hspace{0.5ex}
\includegraphics[width=0.08\columnwidth]{figures/1.pdf}
\hspace{1ex}
{\small joining}
%\hspace{0.5ex}
\includegraphics[width=0.08\columnwidth]{figures/0_5.pdf}
}
\end{center}
\begin{tabular}{cccc}
\hspace{-2ex}\includegraphics[width=0.245\linewidth]{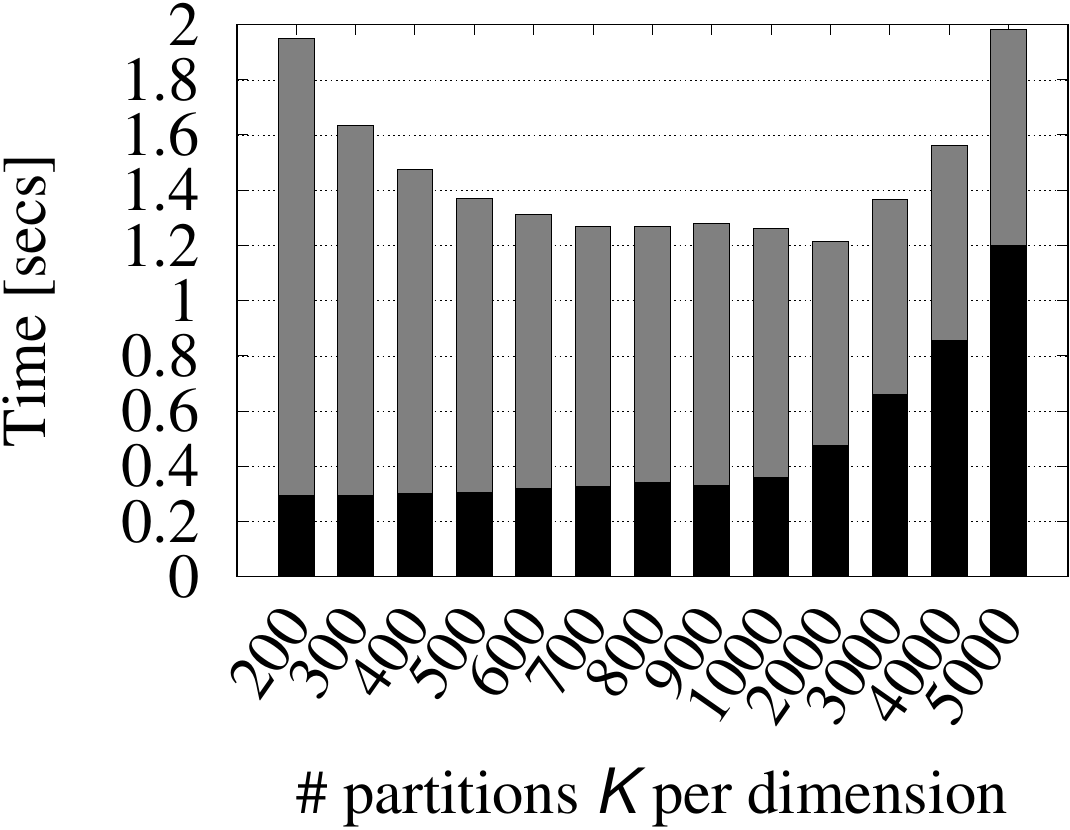}
&\hspace{-1ex}\includegraphics[width=0.245\linewidth]{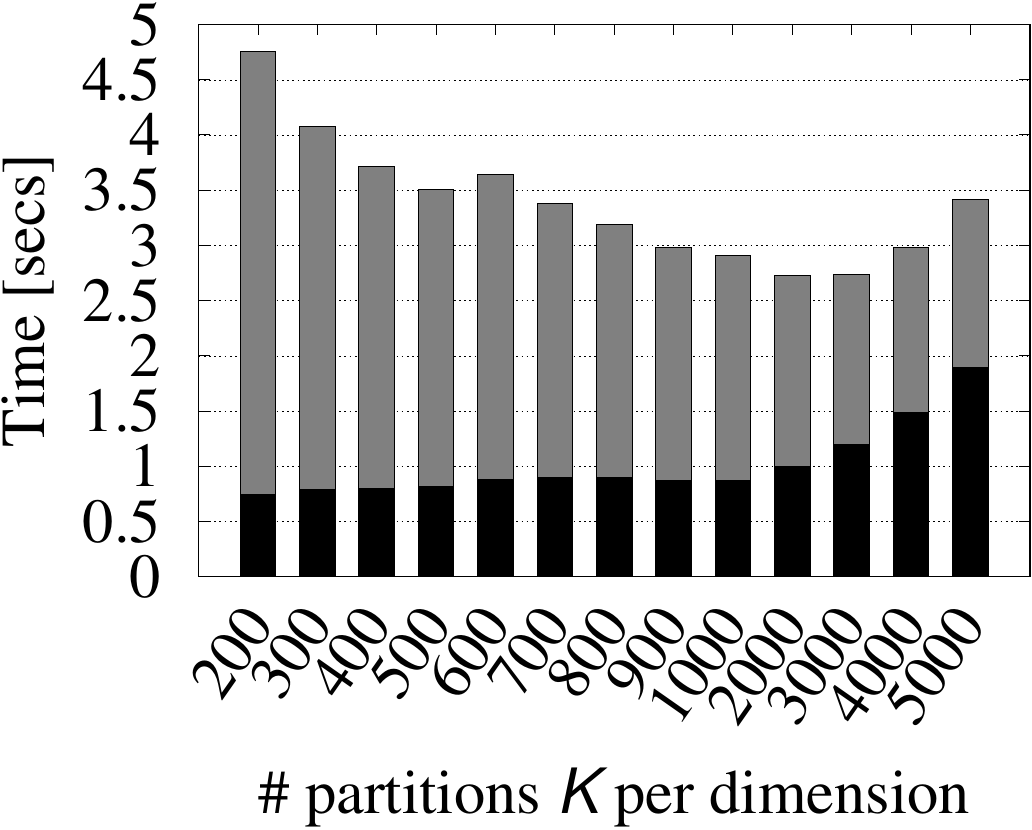}
&\hspace{-1ex}\includegraphics[width=0.245\linewidth]{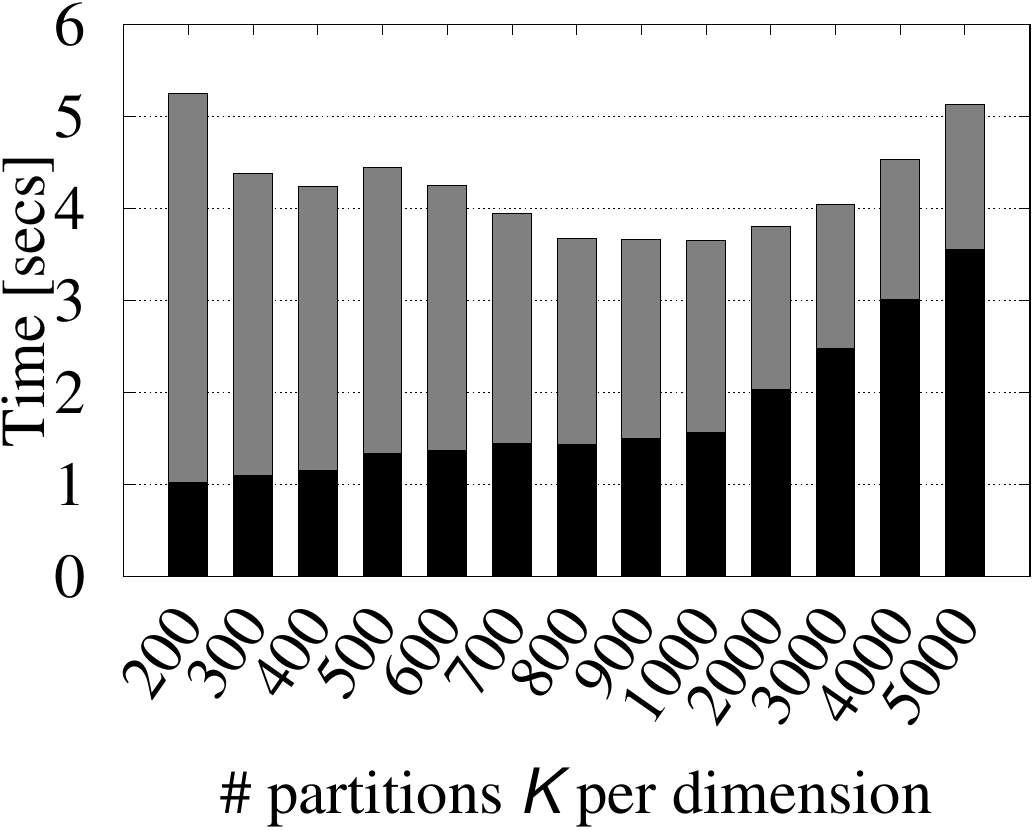}
&\hspace{-1ex}\includegraphics[width=0.245\linewidth]{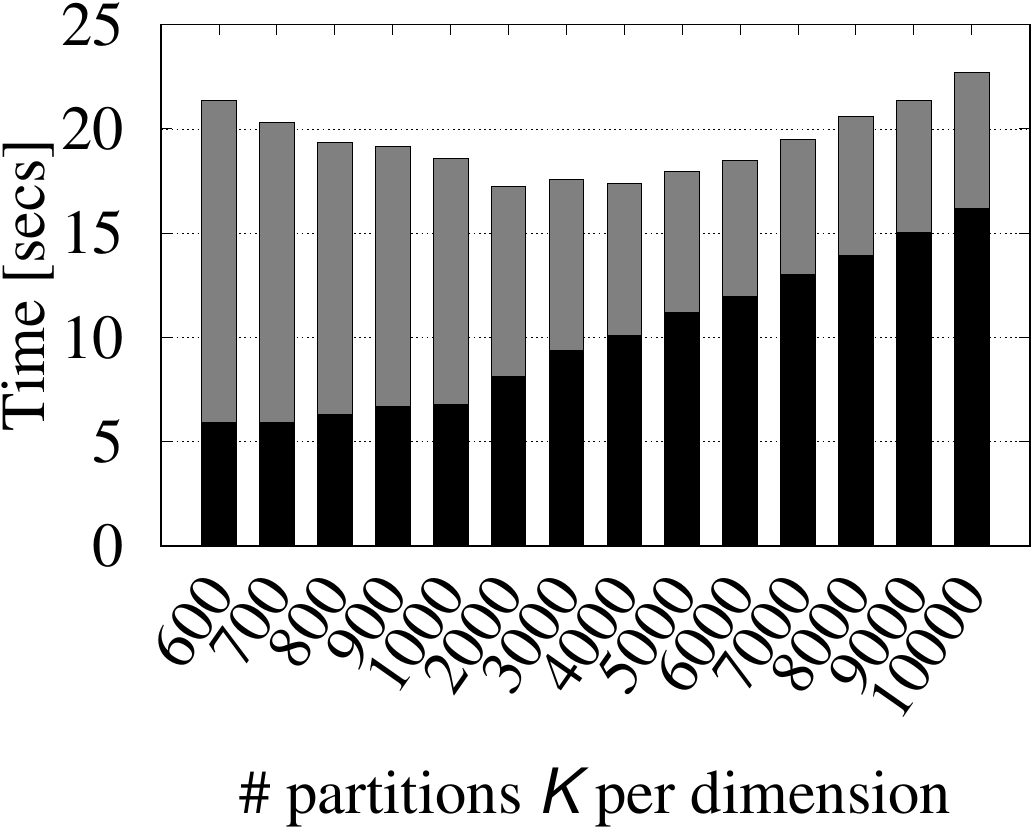}\\
%\scriptsize{\# partitions} &\scriptsize{\# partitions} &\scriptsize{\# partitions} &\scriptsize{\# partitions}\\
(a) $T2 \bowtie T5$ &(b) $T2 \bowtie T8$ &(c) $O5 \bowtie O6$ &(d) $O6 \bowtie O9$\\
\end{tabular}
%\vspace*{-2ex}
\caption{Tuning 2D partitioning: time breakdown}
\label{fig:tuning_2d_breakdown}
%\vspace{-5ex}
\end{figure*}
\subsubsection{Tuning 2D Partitioning}

%
%We now repeat the same set of experiments from the previous section,
%but this time using a 2D partitioning.
We varied the granularity $K\times K$ of the grid and measure
for each value of $K$ the runtime cost of the algorithm, when the
sweeping axis is always set to $x$, always set to $y$, or when our
adaptive model is used to select the sweeping axis at each tile (which
could be different at different tiles).
%Again, we use the GDT approach for handling duplicates.
Figure \ref{fig:tuning_2d} depicts the performance of the three join
variants. The observations regarding the choice of the sweeping axis
and the number of partitions are similar to the cases of 1D
partitioning. Specifically, when the number of partitions is small
$K\le 20$,
the choice of the sweeping axis makes a difference and choosing $x$ is
better. In these configurations, our model can be even better than
always choosing $x$. The three options converge at about $K=500$
and there are no significant differences between them after this
point.

Figure~\ref{fig:tuning_2d_breakdown} shows the cost breakdown for the partitioning
and joining phases of the 2D spatial join, when our model is used for
picking the sweeping axis $x$.
As in the case of 1D joins, we observe that the cost of partitioning
increases with $K$ and becomes too high when the tiles become too many
and very small (i.e., when $K>2,\!000$). On the other hand, the join cost
drops, but stabilizes after $K>2,\!000$. After this point, the number
$K\times K$ of tiles (that have to be managed) becomes significantly high
and replication becomes excessive. The joining phase does not benefit;
due to replication, the join inputs at each tile do not reduce in size
and the same join results are computed in neighboring tiles. To sum up, the best grid configuration is around $K=2,\!000$, which is consistent
with the best option in 1D partitioning. In addition, our tests show that the 2D partitioning version of the algorithm should always use our adaptive model to select the sweeping axis.

\subsubsection{1D vs. 2D Partitioning}
Our experiments in tuning PBSM draw two important observations. First, regarding the number of partitions, the rule of the thumb is to select $K$ (in both 1D and 2D partitioning) such that the
extents of the resulting partitions are about one order of magnitude
larger than the extents of the rectangles (in one or both dimensions, respectively).
For the rest of our analysis, we use this rule to select $K$ as the
default number of divisions in the splitting dimension(s). 

Second, 
%In general 
1D partitioning achieves better performance compared to 2D
partitioning, due to less replication and the fact that all tiles in a
row or a column can be swept by a single line (along the row or
column) with the same effect as processing all tiles independently
with sweeping along the same direction.
Table \ref{tab:speedup} summarizes, for the four join queries, the
best speedups achieved by 1D and 2D partitioning, compared to
the best corresponding performance of the plane sweep algorithm
without partitioning.
1D partitioning is up to 32\% faster compared to 2D
partitioning.

\begin{table}[!t]
\centering
%\caption{Speedup achieved by 1D and 2D partitioning.}
\caption{1D vs. 2D partitioning: speedup}
\label{tab:speedup}
\small
\begin{tabular}{|c|c|c|c|c|}\hline
\multirow{2}{*}{\textbf{query}} &\multicolumn{2}{c|}{\textbf{1D}} &\multicolumn{2}{c|}{\textbf{2D}} \\\cline{2-3}\cline{4-5}
 &$K$ &speedup &$K\times K$
      &speedup\\\hline\hline
$T2 \bowtie T5$ &$3000$		&$9.6$x &$1000\times 1000$ &$8.16$x\\
$T2 \bowtie T8$ &$7000$	& $10.67$x &$2000\times 2000$ &$8.98$x\\
$O5  \bowtie O6$             &$3000$	&$8.62$x &$1000\times 1000$ &$6.82$x\\
$O6 \bowtie O9$          &$7000$ &$16.56$x &$2000\times 2000$ &$12.58$x\\
%T4 $\bowtie$ T8 &$674.50$s &$1,\!360.92$s &$8,\!135$  &$19,\!406$\\
%O9 $\bowtie$ O3 &$926.14$s &$1,\!681.30$s&$4,\!535$ &$11,\!529$\\
%\pbour{we need more queries here...}\\
\hline
\end{tabular}
\end{table}

\begin{table}
\caption{Parallel evaluation: runtime (1D partitioning)}
%\vspace{-2ex}
\centering
\small
\label{tab:par_speedup}
\begin{tabular}{|c|c|c|c|c|}\hline
\multirow{2}{*}{\textbf{\# threads}} &\multicolumn{4}{c|}{\textbf{queries}}\\\cline{2-5}
 &$O5 \bowtie O6$ &$O6 \bowtie O9$ &$T4 \bowtie T8$ &$O9 \bowtie O3$\\\hline\hline
1 &2.98s		&14.4s		&20.1s		&43.0s\\
5 &0.75s		&3.32s		&4.34s		&10.6s\\
10 &0.46s		&1.91s		&2.47s		&6.11s\\
15 &0.38s		&1.45s		&1.85s		&4.54s\\
20 &0.32s		&1.21s		&1.64s		&3.54s\\
25 &0.29s		&1.07s		&1.42s		&3.09s\\
30 &0.28s		&0.99s		&1.36s		&2.89s\\
35 &0.27s		&0.96s		&1.27s		&2.72s\\
40 &0.27s		&0.91s		&1.21s		&2.72s\\\hline
\end{tabular}
\end{table}
\subsection{Parallel Evaluation}
In the last experiment, we tested the parallel version of the
algorithm. We used 1D partitioning which is superior to 2D partitioning as shown
in Table~\ref{tab:speedup}. %--\ref{tab:gbda}.
%Figure \ref{fig:parallel_1d} shows the cost of the algorithm when
%using GDT or GBDA for various pairs of joined datasets as a function
%of the number of threads used. 
Table~\ref{tab:par_speedup} summarizes, for the four join queries, the
runtime and the speedup achieved by our parallel evaluation of the spatial join.
Note that the performance scales
gracefully with the number of threads, until it stabilizes over 20
threads, %where hyper-threading starts having effect.
which equals the number of physical cores in our machine.
%GDT is slightly better than  GBDA in all cases, except for the
%$T4 \bowtie T8$ join where GBDA is better.
As a general conclusion our parallel design takes full advantage of
the system resources to reduce the join cost in the order of a few
seconds. We are not aware of any previous work that can achieve such a
performance when joining datasets in the order of several millions of
objects each. 
%\fix{++ compare join-only cost, assuming that data are already
%  partitioned as in a data management system (e.g. SpatialHadoop)
%  which uses the grid as an index for queries. See if mj could beat
%  ditt in this case.
%}

\section{Conclusions}\label{sec:con}
In this paper, we have investigated directions towards tuning a
classic and popular partitioning-based spatial join algorithm, which
is typically used for in-memory and parallel/distributed join
evaluation. We investigated the tuning of the algorithm by varying the
number and type of partitions\eat{, the method used for duplicate
avoidance,} and the sweeping axis choice in plane sweep. We also
designed an efficient parallel version of the algorithm. Our
experimental findings show that 1D partitioning performs better than
2D and that the correct sweeping axis choice does matter. In addition,
we showed that the parallel version of the algorithm scales well with
the number of threads.  

Directions for future work include consideration of the refinement
step of the join, which can be significantly more expensive than the
filter step. In addition, we plan to adapt our techniques and
investigate their performance in a distributed environment and for the
case of NUMA architectures. 

\balance
%Finally, we plan to study in more detail
%for which inputs GBDA performs better than GDT.  
\bibliographystyle{ACM-Reference-Format}
\bibliography{sjoin}

\end{document}